\documentclass[12pt,document,nofootinbib,superscriptaddress, onecolumn,preprintnumbers,balancelastpage]{revtex4}
\pdfoutput=1
\usepackage{latexsym}
\usepackage{amssymb}
\usepackage{epsfig,amsmath,graphicx}
\usepackage{listings}
\usepackage{color}
\usepackage{dcolumn}
\usepackage{slashed}
\usepackage{comment,latexsym} 
\usepackage{bm}
\usepackage{verbatim}
\usepackage{longtable,tabularx}
\usepackage{dcolumn}
\usepackage{chngpage}
\definecolor{Mahogany}{rgb}{0.62,0.24,0.15}
\definecolor{colorLink}{rgb}{0.7,0,0}
\definecolor{colorCite}{rgb}{0,.7,0}
\definecolor{colorURL}{rgb}{0,0,0.7}
\usepackage[colorlinks=true,linktocpage=true,linkcolor=black,citecolor=colorCite,urlcolor=colorURL]{hyperref}
\usepackage{setspace}
\usepackage{xspace}
\usepackage{multirow}
\usepackage{hhline}
\usepackage{titlesec}
\usepackage{comment}

\newcommand\T{\rule{0pt}{2.6ex}}
\newcommand\B{\rule[-1.2ex]{0pt}{0pt}}

\newcommand{\LL}{\mathcal{L}}

\def\be{\begin{equation}}
\def\ee{\end{equation}}
\newcommand{\beq}{\begin{equation}}
\newcommand{\eeq}{\end{equation}}
\def\bea{\begin{eqnarray}}
\def\eea{\end{eqnarray}}

\newcommand{\GeV}{{\text{GeV}}}
\newcommand{\TeV}{{\text{TeV}}}

\newcommand{\ifb}{{{\text{ fb}}^{-1}}}

\newcommand{\MET}{\slashed{E}_T}

\newcommand{\scinot}[2]{#1\times 10^{#2}}

\newcommand{\ts}{\widetilde{t}}
\newcommand{\X}{\widetilde{\chi}^0_1}
\newcommand{\sys}{\varepsilon_{\textrm{sys}}}

\expandafter\def\expandafter\normalsize\expandafter{%
    \normalsize
    \setlength\abovedisplayskip{8pt}
    \setlength\belowdisplayskip{8pt}
    \setlength\abovedisplayshortskip{8pt}
    \setlength\belowdisplayshortskip{8pt}
}

\titleformat{\section}{\center\normalfont\fontsize{14}{15}\bfseries}{\thesection.}{1em}{}
\titleformat{\subsubsection}{\center\normalfont\fontsize{12}{15}}{\thesubsubsection.}{1em}{}

\begin{document}

\begin{flushright}
\text{\normalsize SLAC-PUB-15987}
\end{flushright}
\vskip 15 pt

\title{Boosting Stop Searches with a 100 TeV Proton Collider} 

\author{$\quad\quad\quad\quad\quad\quad\quad\quad$ Timothy Cohen}
\affiliation{
Theory Group, SLAC National Accelerator Laboratory\\
\vskip -6 pt
Menlo Park, CA 94025}

\author{Raffaele Tito D'Agnolo}
\affiliation{
School of Natural Sciences, Institute for Advanced Study\\
\vskip -6 pt
Princeton, NJ 08540}

\author{$\quad\quad\quad\quad\quad\quad\quad\quad$ Mike Hance}
\affiliation{
Lawrence Berkeley National Laboratory\\
\vskip -6 pt
Berkeley, CA 94720}

\author{Hou Keong Lou}
\affiliation{
Physics Department, Princeton University\\
\vskip -6 pt
Princeton, NJ 08544}

\author{Jay G. Wacker\,}
\affiliation{
Quora\\
\vskip -6 pt
Mountain View, CA 94041}
\affiliation{
Stanford Institute for Theoretical Physics\\\vskip -6 pt
Physics Department, Stanford University\\\vskip -6 pt
Stanford, CA 94305}

\begin{abstract}
\vskip 5 pt
\begin{center}
{\bf Abstract}
\end{center}
\vskip -30 pt
$\quad$
\begin{spacing}{1.1}
A proton-proton collider with center of mass energy around 100 TeV is the energy frontier machine that is likely to succeed the LHC.  One of the primary physics goals will be the continued exploration of weak scale naturalness.  Here we focus on the pair-production of stops that decay to a top and a neutralino.  Most of the heavy stop parameter space results in highly boosted tops, populating kinematic regimes inaccessible at the LHC.  New strategies for boosted top-tagging are needed and a simple, detector-independent tagger can be constructed by requiring a muon inside a jet.  Assuming 20\% systematic uncertainties, this future collider can discover (exclude) stops with masses up to 5.5 (8) TeV with $3000\ifb$ of integrated luminosity.  Studying how the exclusion limits scale with luminosity motivates going beyond this benchmark in order to saturate the discovery potential of the machine.
\end{spacing}
\end{abstract}

\maketitle
\newpage
\begin{spacing}{1.3}

\section{Introduction}
\label{sec:introduction}
Exploring the nature of our Universe at the smallest possible scales is the primary goal of the particle physics community.  This pursuit will require extending the energy frontier program beyond the 14 TeV LHC.

Recently the idea of building a 100 TeV circular proton-proton collider has gained momentum, starting with an endorsement in the Snowmass Energy Frontier report \cite{Gershtein:2013iqa}, and importantly followed by the creation of two parallel initiatives: one at CERN~\cite{FCC} and one in China~\cite{SppC}.  For some recent studies of the capabilities of a 100 TeV collider see \cite{Cohen:2013xda, Andeen:2013zca, Apanasevich:2013cta, Degrande:2013yda, Stolarski:2013msa, Yu:2013wta, Zhou:2013raa, Low:2014cba}.  Regardless of what is discovered during the upcoming run of the LHC, data from the 100 TeV machine will still be utilized to push new particle searches to higher mass scales.  The existence (or absence) of these states could have a dramatic impact on the way we think about fundamental questions.  Of particular interest to this work is the question of weak scale naturalness, and specifically the possibility of TeV-scale top partners.  Any discoveries of such particles at the LHC would likely require further study at a higher-energy machine.
However, even in the event that the LHC does not find any top partners, this program will continue to be of central importance by pushing the fine-tuning of the Higgs mass into qualitatively new regimes.

Here we focus on the stop in supersymmetric (SUSY) extensions of the Standard Model (for a SUSY status update after Run I of the LHC, see \cite{Craig:2013cxa}). Naturalness considerations imply that the stops should be light \cite{Dimopoulos:1995mi, Cohen:1996vb} in order to regulate the Higgs mass, while the masses of first and second generation squarks are less constrained.  Explicit models that realize the so-called ``natural SUSY" spectrum have been constructed \cite{ArkaniHamed:1997fq, Craig:2011yk, Csaki:2012fh, Larsen:2012rq, Craig:2012hc, Cohen:2012rm, Randall:2012dm}, and often the dominant collider signatures can be reduced to a set of now-standard Simplified Models~\cite{Alves:2011wf} involving only the third generation squarks, a neutralino, and a gluino \cite{Essig:2011qg, Papucci:2011wy, Brust:2011tb}. A discovery of stops, or at least an understanding of the allowed parameter space of these models, has direct implications for weak-scale naturalness.

We study the stop-neutralino Simplified Model, in which the stops are pair-produced, and each stop decays to a top and a stable neutralino\footnote{The minimal natural spectrum in the 
MSSM 
is slightly more complicated, due to the expectation that both stops, the left-handed sbottom, and the Higgsinos will all be light.  The model studied here provides similar reach for the majority of the parameter space of these more complete models \cite{Kribs:2013lua, Papucci:2014rja}.}.  This signature is well suited to compare the physics implications of different machine parameters such as $\sqrt{s}$ and total integrated luminosity.

Searches for direct stop production have been carried out at both ATLAS and CMS, providing limits on the stop mass of $\approx$ 800 GeV with 20 fb$^{-1}$ at $\sqrt{s}=8$ TeV \cite{ATLAS-CONF-2013-024, ATLAS-CONF-2013-037, ATLAS-CONF-2013-065, Chatrchyan:2013xna, CMS-PAS-SUS-13-004}.  The high-luminosity upgrade of the LHC (HL-LHC) is expected to deliver $3000\ifb$ of data at $\sqrt{s}=14$ TeV, allowing for a discovery reach of $\approx$ 800 GeV stops and an exclusion reach of $\approx$ 1.5 TeV \cite{ATLAS:2013hta, CMS:2013xfa}.  

Beyond naturalness considerations, this study is motivated by the exploration of new kinematic regimes in top physics. In 100 TeV collisions, the tops from stop decays are so highly boosted that current LHC analysis strategies, usually based on resolving the individual decay products of the top, become ineffective.  This work demonstrates that an analysis that relies on a muon inside a jet can be used to discover (exclude) stop masses up to $\approx 5.5\, (8) $ TeV.

One issue in the specifications of the 100~TeV collider that has not yet been addressed is the integrated luminosity needed to fulfill its physics potential.  The baseline integrated luminosity is taken to be $3000\ifb$, but we also consider scenarios yielding $300\ifb$ and $30000\ifb$.  We find that $3000\ifb$ may be insufficient to saturate the physics reach of a high-energy machine. 

The rest of this paper is organized as follows.  Section~\ref{sec:sub} studies generic properties of heavy new physics decaying to boosted tops and compares the sensitivity of jet substructure techniques and muon-in-jet requirements. Section~\ref{sec:strategy} presents a cut flow optimized for heavy stops that is based on the presence of a muon inside a jet and shows its sensitivity in the stop-neutralino mass plane; an additional analysis is presented which optimizes the reach for compressed spectra. Section~\ref{sec:conclusion} summarizes the implications of the analysis on future accelerator and detector design, and discusses the implications of the mass sensitivity for fine-tuning.  

The results presented here rely on events generated at parton level with \texttt{MadGraph5}~\cite{Alwall:2011uj}, showered with \texttt{Pythia6}~\cite{Sjostrand:2006za}, and processed using \texttt{Delphes}~\cite{deFavereau:2013fsa} and the Snowmass combined detector card \cite{Anderson:2013kxz}.  The stop signals are normalized to the NLL + NLO cross sections computed in \cite{Borschensky:2014cia}. The Snowmass background samples \cite{Avetisyan:2013onh} were used, augmented by a high statistics $H_T$-binned QCD sample generated for this study. 

\section{Boosted Tops at 100 TeV}\label{sec:sub}
Signal events in the stop-neutralino Simplified Model include pair-produced stops ($\widetilde t$) that decay promptly into a top quark and a stable neutralino ($\X$).  
Under the assumption that the stops are produced at rest, the boost of the top quark is given by
\be
\gamma_{t} = 
\frac{m_{\ts}}{2\,m_{t}}\left(1- \frac{m_{\X}^2-m_t^2}{m_{\ts}^2}  \right)
\label{eq:top_boost}
\ee
and the resulting top jet has a typical size of $\Delta R\sim 1/\gamma_t\sim m_t / p_T^t$.

The left panel of of Fig.~\ref{fig:topboost} shows the $p_{T}$ distribution of the leading top quark for three different stop masses (assuming a massless neutralino).  For stops with a mass of a few TeV or higher, the tops from the stop decay are highly boosted with $p_{T}\gg m_t$. The right panel of Fig.~\ref{fig:topboost} shows the mean distance between the $W$ boson and the $b$ from the decay of the top as a function of $m_{\ts}$ and $m_{\X}$.

Given that the jet radius chosen for this study is $\Delta R = 0.5$, the top will on average be contained within a single jet.  Stop searches at a 100 TeV collider will therefore have to probe a kinematic regime not accessible to the 14 TeV LHC, where the top $p_T$ relevant for most searches is less than a TeV.

\begin{figure}[ht]
\raisebox{-0.048\height}
{\includegraphics[trim=3mm 0 0 0,clip=true, width=.45 \textwidth]
{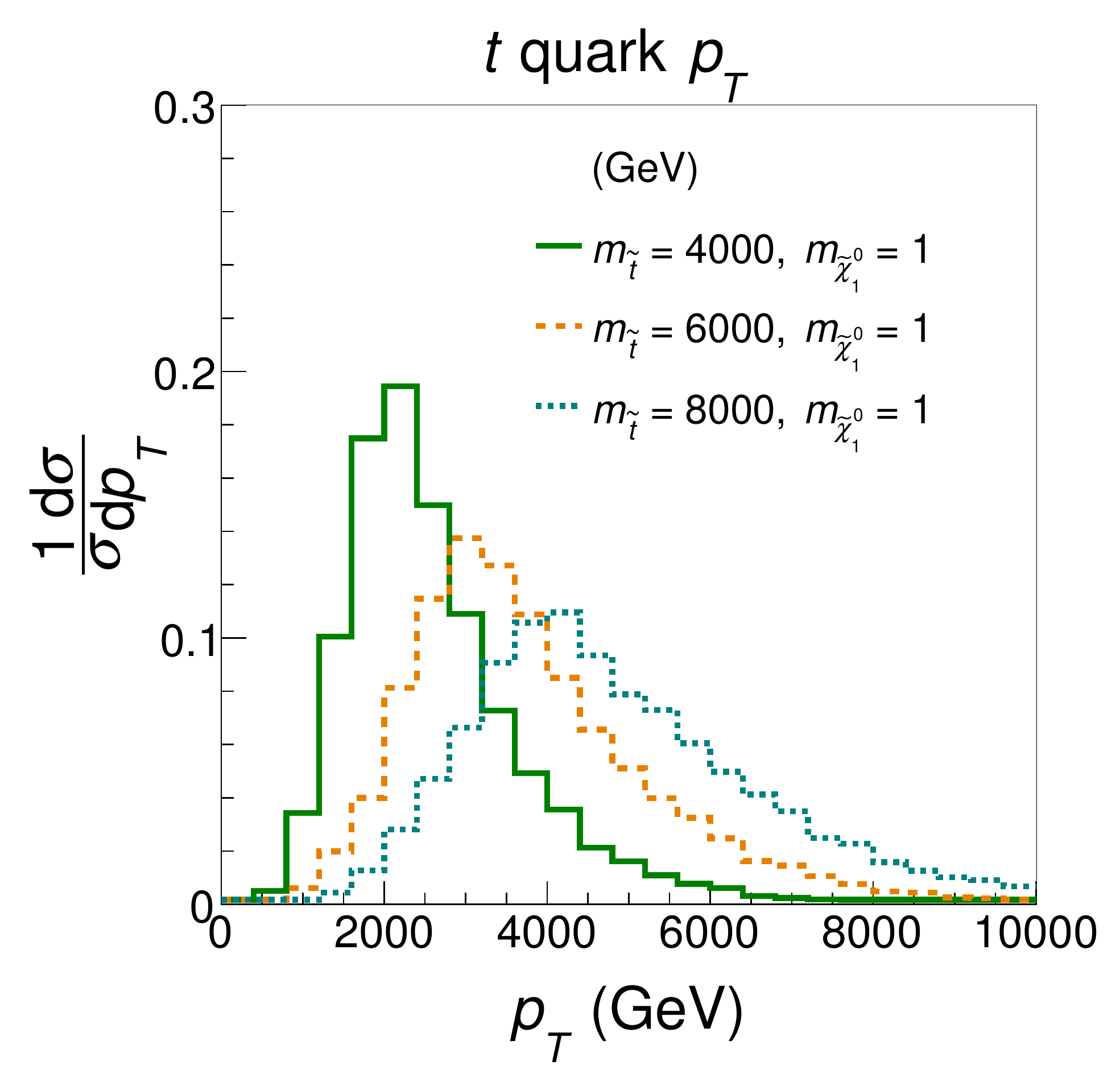}}
\includegraphics[width=.407 \textwidth]{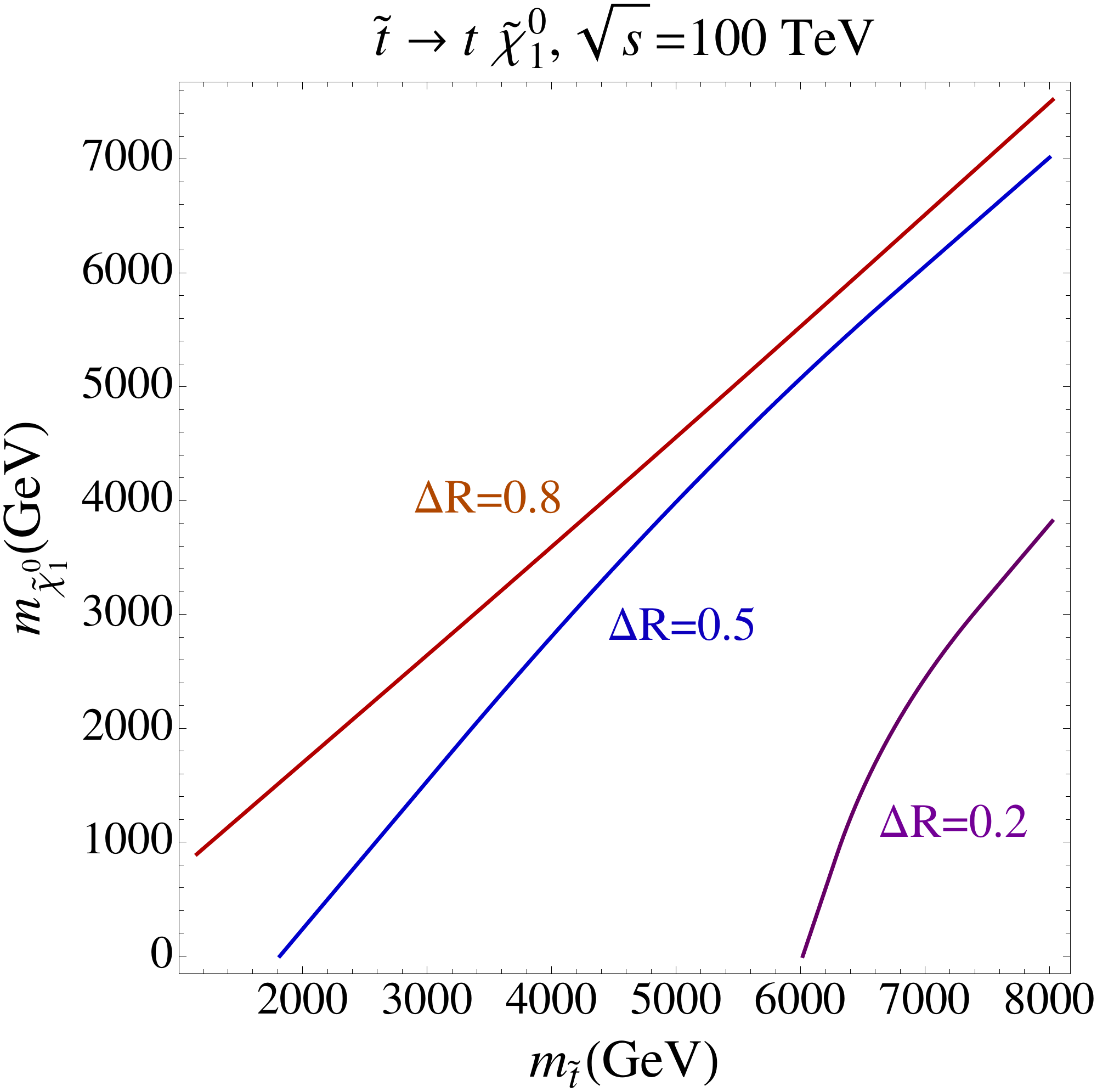}
\caption{The $p_T$ distribution of the leading top quark for $m_{\ts}=2,6,10$ TeV assuming $m_{\X} = 1\, \GeV$ [left].  The average size of top jets from stop decays as a function of $m_{\ts}$ and $m_{\X}$ [right].  }
\label{fig:topboost}
\end{figure}


One possible tool for separating signal from background is to tag these highly boosted tops.  Note that top taggers constructed for LHC energies are optimized for large radius jets with $\Delta R \approx 1.0 - 1.5$ (for a review, see \cite{Plehn:2011tg}).  It is therefore interesting to understand if existing algorithms are suitable for events at 100 TeV.  If the top tagger depends on an intrinsic angular scale, for example the Johns Hopkins top tagger~\cite{Kaplan:2008ie}, then the choices appropriate for tagging boosted tops at the LHC will need to be reconsidered.  In contrast, the HEP top tagger~\cite{Plehn:2010st} does not make any assumption about the angular separation of the top decay products.

Given the magnitude of the boost being considered, separating the individual constituents of the top decay requires detector granularities higher than presently available in hadronic calorimeters.  For example, a 5 TeV top jet falls within a cone size of roughly $\Delta R\approx 0.07$, while the typical size of a calorimeter cell at ATLAS is $\Delta \eta \times \Delta \phi \sim \mathcal{O}\big(0.1\times 0.1 \big)$~\cite{ATLAS:1999uwa}.  
In order to understand this effect quantitatively, we generated a sample of $t\,\overline{t}$ and QCD events at $\sqrt{s} = 100 \;\TeV$, with a minimum generator-level $p_T$ cut on the leading top at 500 GeV or 5000 GeV.  The hadron-level events were passed through a \texttt{FASTJET}~\cite{Cacciari:2011ma, Cacciari:2005hq} based code.  This framework was validated against the results in \cite{Plehn:2010st} using a sample of 14 TeV events. In order to investigate the impact of finite calorimeter resolution, a simple pixelation was applied by summing particle energies within square cells whose widths were allowed to vary from 0 to 0.1.  The events were then clustered using the Cambridge/Aachen algorithm~\cite{Dokshitzer:1997in, Wobisch:1998wt}, where $\Delta R = 1.5\, (0.5)$ was taken for $p_T > 500 \, (5000)\, \GeV$.  The HEP top tagger was applied to the leading two jets in order to determine the efficiency for tagging a top jet and the probability of mis-tagging a QCD jet.  The results are shown in Fig.~\ref{fig:top_tagging}, where top-tagging is found to be insensitive to the detector granularity for 500 GeV top jets, but with a cell width $\gtrsim 0.02$ is significantly degraded for 5 TeV top jets.  

The jet radius changes approximately as the inverse of the top $p_T$, $\Delta R \sim m_t/p_T^t$, so in most of the parameter space of interest for this simplified model, this toy study demonstrates that a much finer calorimeter segmentation than that used for LHC detectors will be needed to exploit substructure techniques at higher-energy colliders.  On the other hand, tracking systems have much finer granularity than is needed by the HEP top tagger, so it would be interesting to explore a Particle Flow or a purely track-based approach.  We leave this for future studies.

\begin{figure}[ht]
\includegraphics[width=.45 \textwidth]{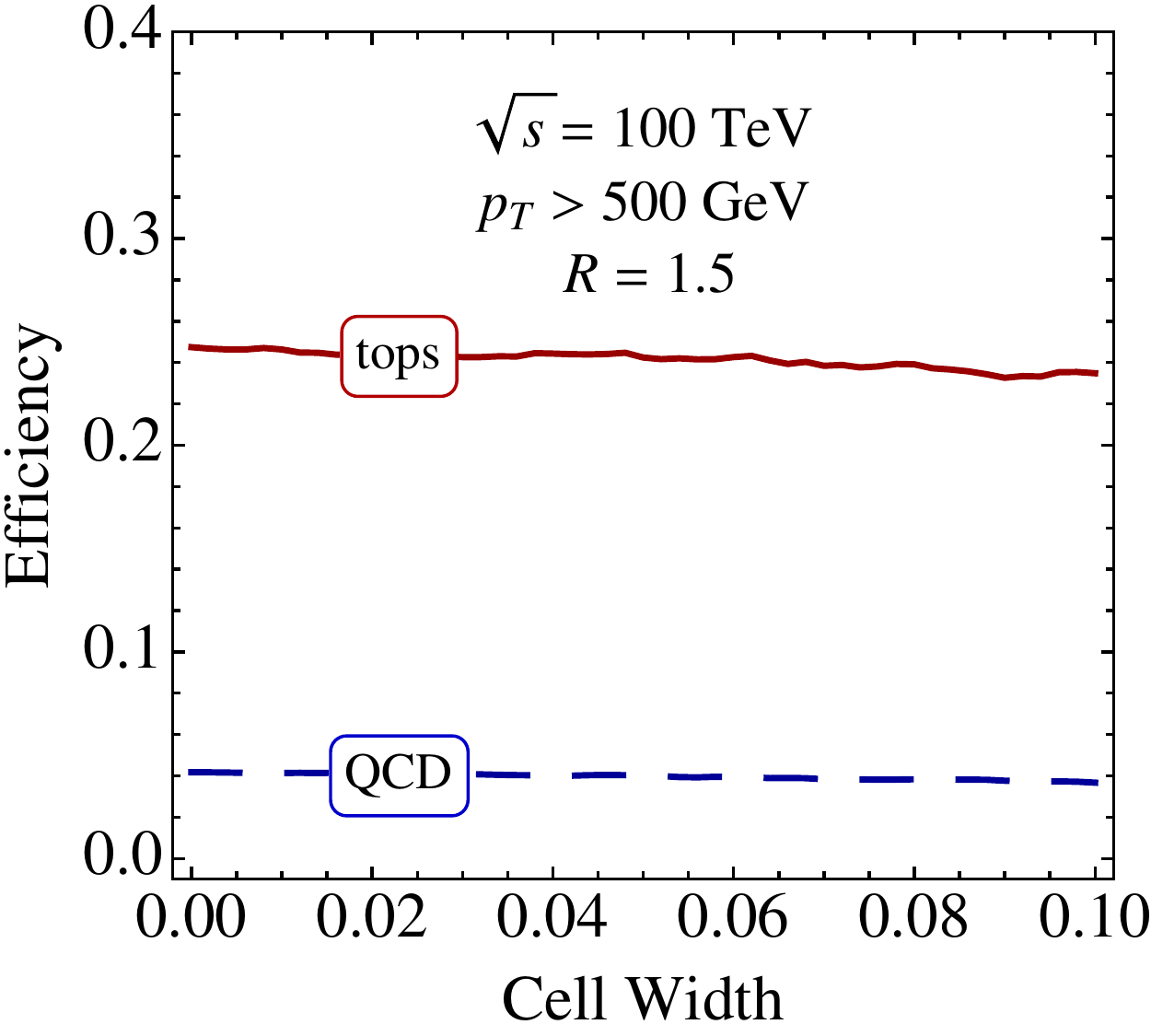} $\quad$ \includegraphics[width=.45 \textwidth]{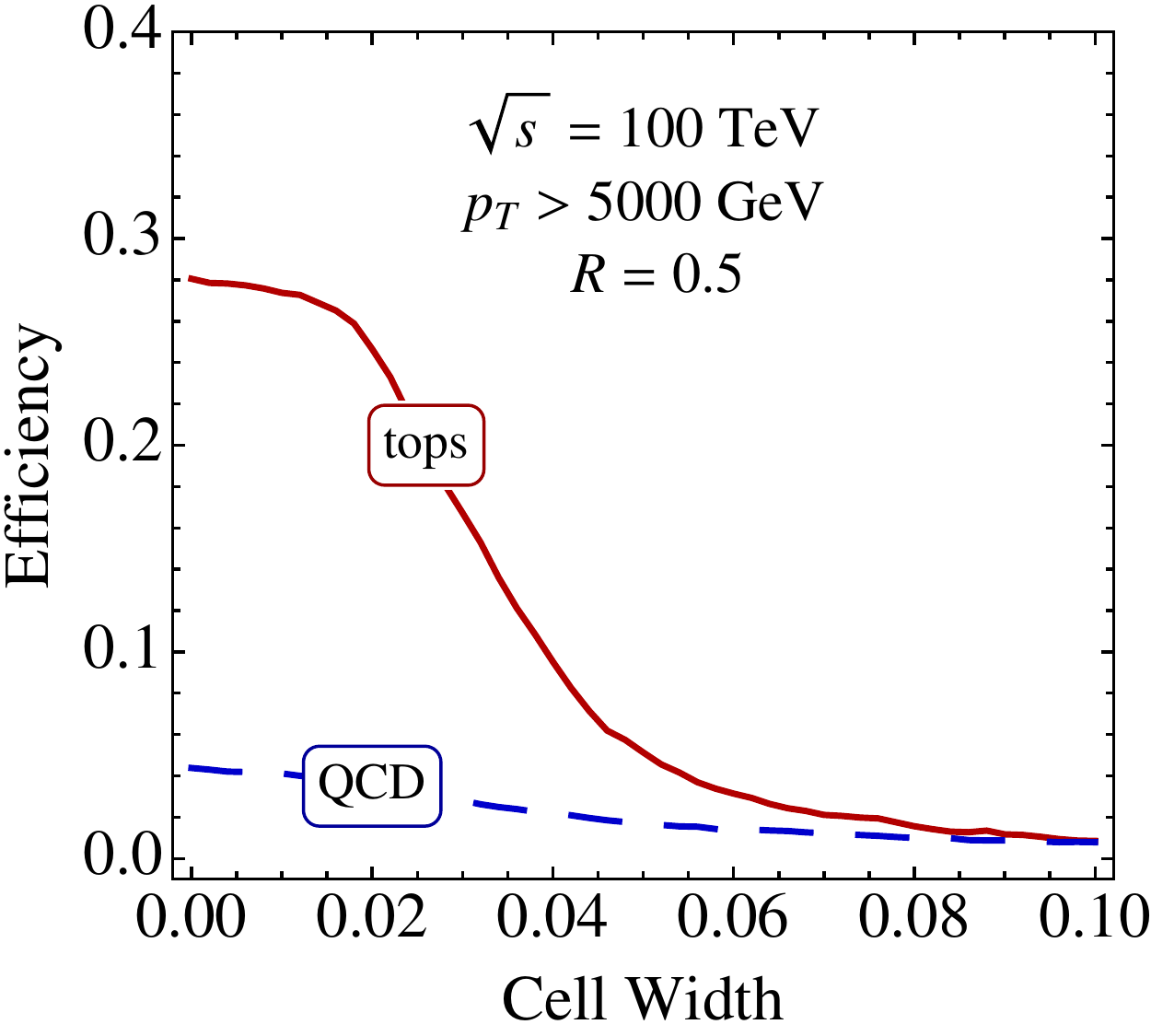}
\caption{HEP top tagger performance for jets with $p_{T} > $ 500 GeV [left] and $ > 5$ TeV [right]. The red solid curve shows the tagging efficiency for top quarks, and the blue dashed curve shows the mis-tag rate for light-flavor QCD jets.}
\label{fig:top_tagging}
\end{figure}

Instead, we consider an alternative strategy with less sensitivity to the detector response.  When a highly-boosted top decays leptonically, or when the resulting $b$ (or even $c$) quark decay yields a lepton, it is very likely that the lepton(s) will be collinear with the top jet.  Requiring a hard lepton inside a jet can therefore be used to tag boosted tops \cite{Thaler:2008ju}. Tagging a top jet by a muon is similar to leptonic $b$-tagging techniques implemented at the Tevatron \cite{Acosta:2005zd, Abazov:2004bv, Abulencia:2005qa, Aaltonen:2007dm, Aaltonen:2009ad, Aaltonen:2010se} and at the LHC~\cite{Aad:2009wy, ATLAS:2010afa, Chatrchyan:2012jua, ATLAS-CONF-2013-068}.  By definition these leptons will not be isolated from nearby tracks or calorimeter activity, removing a common handle for rejecting fake leptons.  For simplicity we therefore only consider the case where a muon is collinear with a jet, and assume that a layered detector design similar to that employed by LHC experiments will provide adequate rejection of fake muons.  Rejection of fake electrons without the use of an isolation requirement is more detector-dependent, and is not considered here.

Figure~\ref{fig:muon_efficiency} shows the probability of finding a 200 GeV muon within a $\Delta R < 0.5$ cone of the leading jet as a function of the leading jet $p_{T}$ for several signal and background samples.\footnote{Due to the structure of the Snowmass detector card, we are using generator level muons when computing the muon-in-jet requirement.  This procedure was validated against a dedicated sample that was produced with no lepton isolation requirements imposed, thereby giving detector level muons inside jets.} 
The signal efficiency for this requirement is roughly 15\%, compared to $t\,\overline{t} + W/Z$ efficiencies of 3\%, $t\,\overline{t}$ efficiencies of about 2\% and QCD efficiencies around 0.4\%.

For the $t\,\overline{t}$ background, the top quarks constitute only $\sim$ 60\% to the total jet $p_{T}$ in the highly boosted regime, indicating a significant contribution from additional QCD radiation.  This leads to a lower efficiency for $t\,\overline{t}$ than in signal events, where more of the total $p_{T}$ is carried by top jets.

\begin{figure}[!t]
\includegraphics[width=.5 \textwidth]{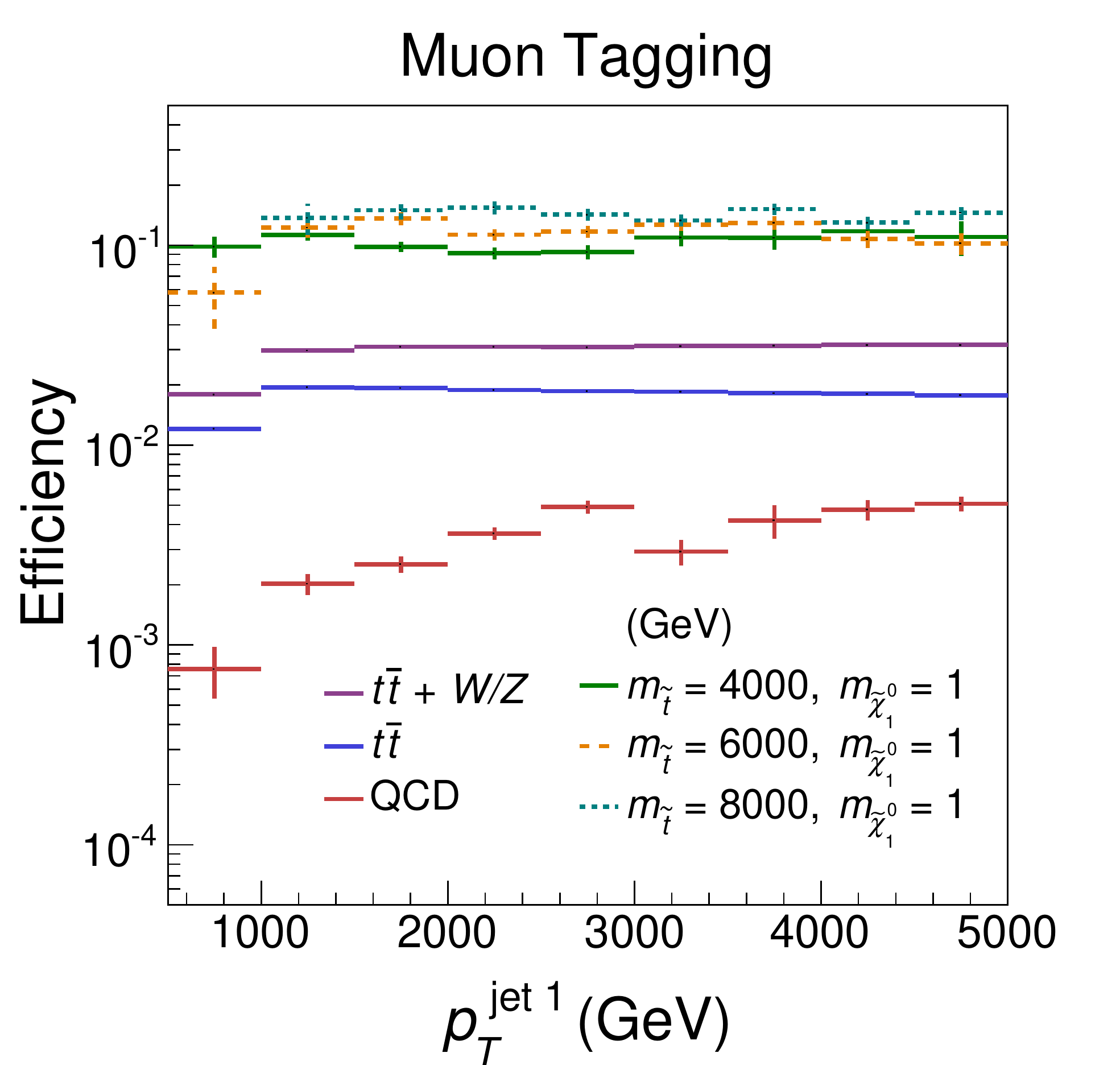}
\caption{Efficiency for finding a $\mu^\pm$ with $p_{T} > 200$ GeV within $\Delta R < 0.5$ of the leading jet for three choices of stop mass, along with the $t\,\overline{t}+W/Z$, $t\,\overline{t}$ and QCD backgrounds.  
}
\label{fig:muon_efficiency}
\end{figure}

Our results in this section ignore the impact of any additional $p\,p$ interactions (pile-up) in the event.  However, we expect pile-up would only degrade the performances of hadronic taggers compared to the muon-in-jet requirement.  Furthermore, it has been shown~\cite{Barletta:2013ooa, Skands:2013asa} that minimum bias events do not change dramatically going from $\sim10$~TeV to $\sim100$~TeV. A larger difference between the LHC and this machine will arise from changes in instantaneous luminosity and/or bunch spacing. If they are within a factor of a few of those at the LHC, it is not inconceivable that a Particle Flow based subtraction scheme could make the performances of substructure techniques (almost) pile-up independent at a 100 TeV collider.

\section{Search Strategy and Results}
\label{sec:strategy}
In the previous section we discussed some general aspects of searches with boosted tops at 100 TeV.  Here we propose a detailed analysis strategy that utilizes the muon-in-jet requirement, and we show the expected discovery reach and 95\% C.L. exclusion sensitivity.  In addition, we provide an alternative cut-flow that relies on isolated leptons in order to increase sensitivity in the compressed region where $m_{\ts} - m_{\X} \approx m_t$.
\subsection{Heavy Stops and Light Neutralinos}
\label{sec:heavy_stop}

We make the following requirements:

\end{spacing}
\vspace{-2pt}
\begin{spacing}{1.15}
\begin{enumerate}
\item At least two anti-$k_T$ jets~\cite{Cacciari:2008gp} with cone parameter $\Delta R=0.5$ and kinematic cuts: $|\eta|<2.5$ and $p_{T}>1000$ GeV. 
\vspace{-5pt}
\item At least one muon with $p_{T\mu}>200$ GeV contained within a $\Delta R=0.5$ cone centered around one of the leading two jets.
\vspace{-5pt}
\item Events with at least one isolated lepton with $p_T > 35$ GeV and $|\eta| < 2.5$ are rejected. The isolation criterion demands the total $p_T$ of all particles within a $\Delta R < 0.5$ cone around the lepton to be less than 10\% of its $p_{T}$.\vspace{-5pt} 
\item $\Delta \phi_{\MET J} > 1.0$, where $\Delta \phi_{\MET J}$ is the smallest $|\Delta \phi|$ between $\MET$ and any jet with $p_{T}>200$ GeV and $|\eta|<2.5$.
\vspace{-5pt}
\item $\MET >$ 3, 3.5 or 4 TeV. Out of the three choices, the cut is chosen for each mass point by optimizing the expected exclusion.
\end{enumerate}
\end{spacing}
\begin{spacing}{1.3}

After imposing a muon-in-jet requirement on the background, the selected sample is composed mainly of boosted heavy quarks. The neutrinos and muons resulting from their decays will be highly collimated and the total $\MET$ will tend to be aligned with the jet momenta.  Therefore it is useful to impose an angular $\Delta \phi$ cut between the $\MET$ and all the jets. For $q\, \overline{q}$, the maximum angle between each neutrino and the final $q$ jet will be of order $m_q/p_{T}$. 
After a stringent $\Delta \phi$ cut, the remaining background is then boosted $t\,\overline{t} + X$ events.  In particular, $t\,\overline{t} + W/Z$ is the dominant background in the signal region. 

\begin{figure}[t!]
\includegraphics[width=.45 \textwidth]{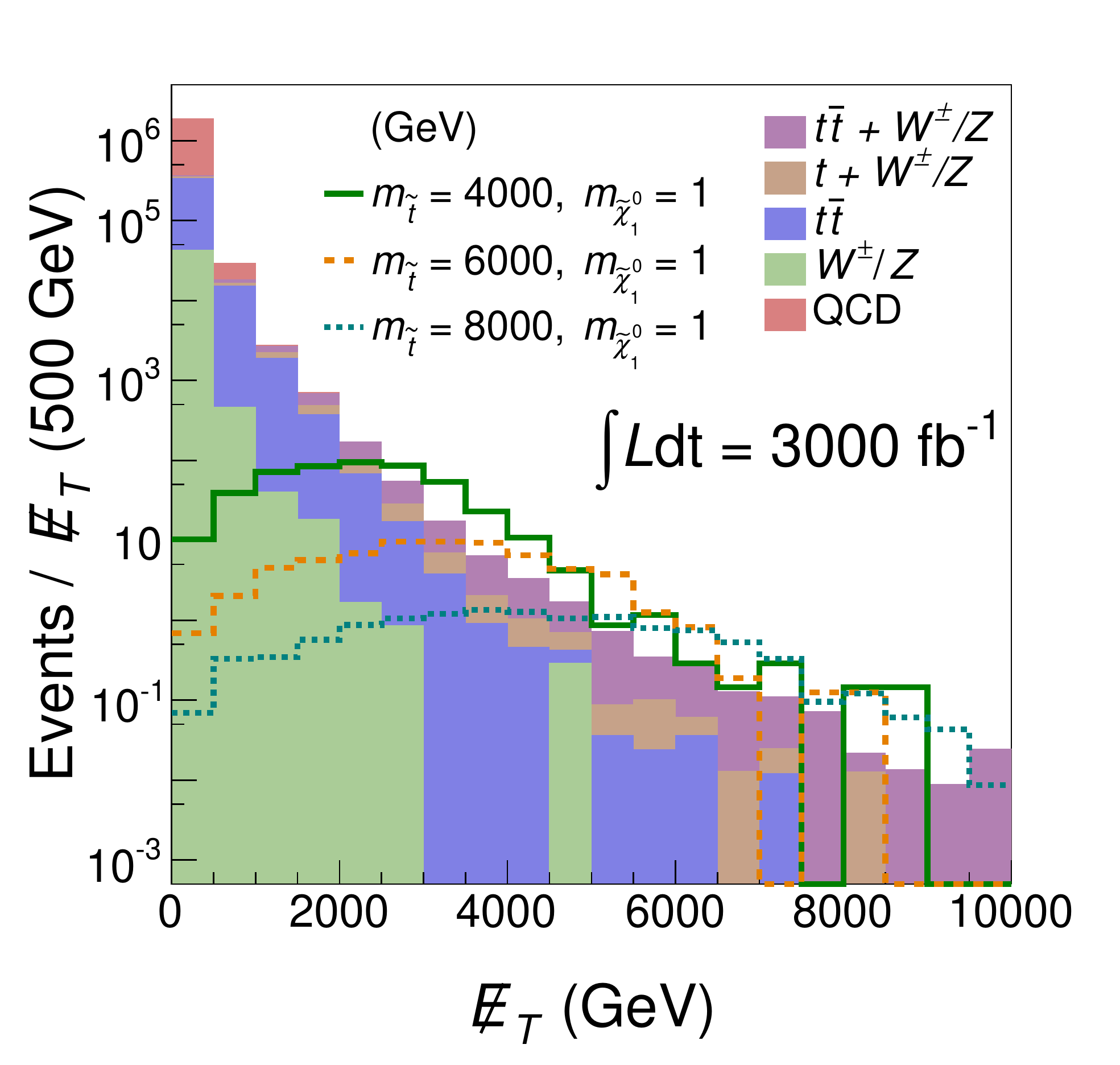}
\includegraphics[width=.45 \textwidth]{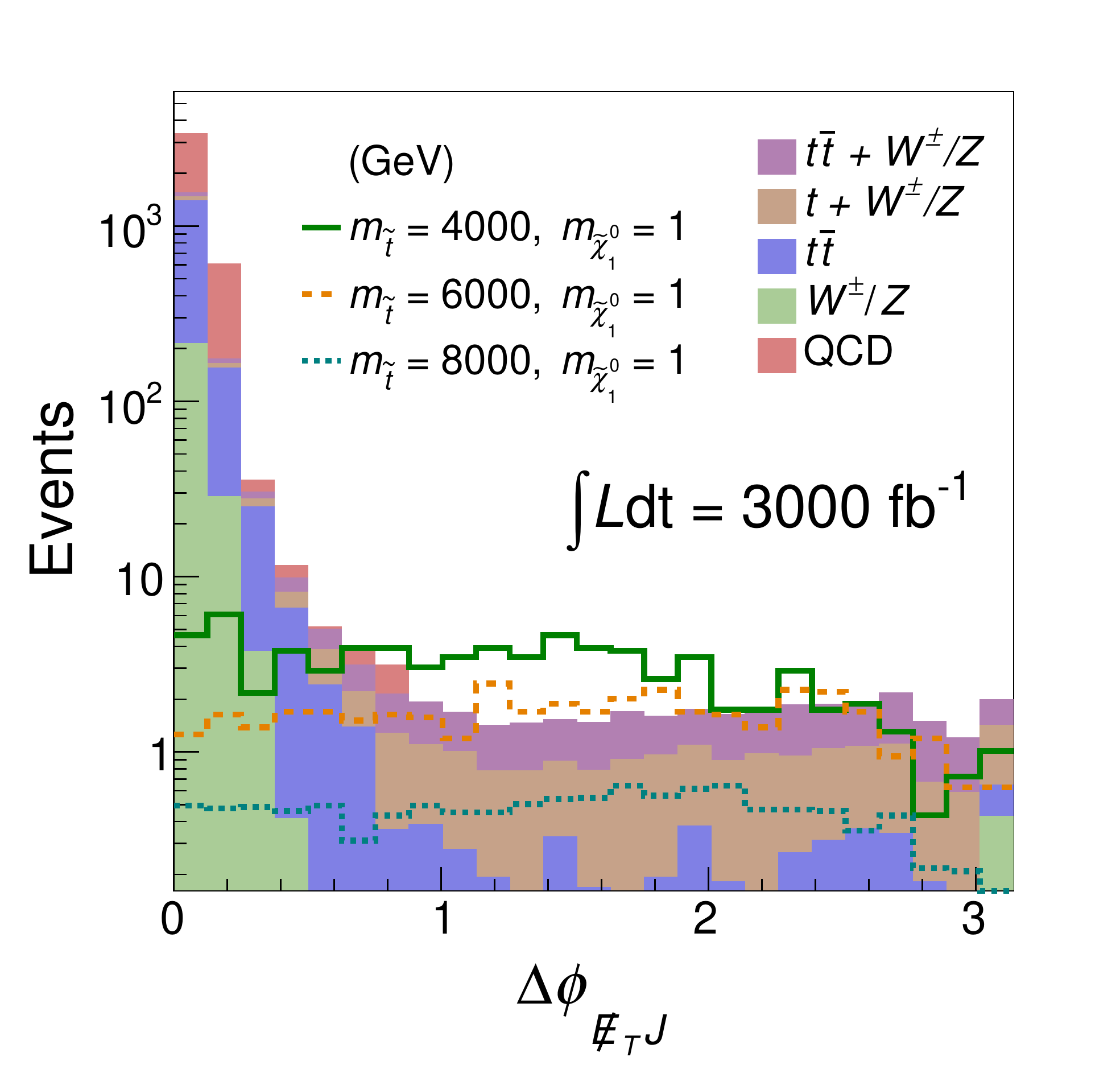}
\caption{The $\MET$ [left] and $\Delta\phi_{\MET J}$ [right] distributions after all other cuts described in Section~\ref{sec:heavy_stop} have been applied, for 3000 fb$^{-1}$ of integrated luminosity.}
\label{fig:met_allbut}
\end{figure}

The $\MET$ and $\Delta \phi_{\MET J}$ distributions after all other cuts are applied are shown in Fig.~\ref{fig:met_allbut}. The low $\MET$ region is mostly dominated by QCD, whereas the high $\MET$ tail is dominated by $t\, \overline t +Z$ ($Z \rightarrow \nu\,\overline{\nu}$).

The results of the cut-flow with $\MET > 4$ TeV for the background and three signal mass points are shown in Table~\ref{table:cut-flow1} without uncertainties.
We note that corrections for electroweak radiation of $W$ or $Z$ bosons within high-$p_T$ jets (e.g. in QCD dijets) could lead to muon-in-jet signatures, and at a high-energy 
machine these corrections can be large~\cite{Christiansen:2014kba}.  We expect that the $\Delta \phi$ requirement will highly suppress such contributions as it already does in events where the $W$ or $Z$ is produced in the matrix element, but this should be verified in future studies.

As a baseline, we choose the relative background and signal uncertainty to be 20\%, and an integrated luminosity of $3000\ifb$.  However, it is useful to explore the reach of this future collider for different choices of systematics and integrated luminosities, especially to study the impact of potential accelerator and detector designs.  In Sec.~\ref{sec:lumi} we therefore show results for a range of integrated luminosities (with appropriate adjustments to the final $\MET$ cut for optimization).  In Sec.~\ref{sec:sys}, we present discovery and exclusion reaches for different choices of systematic uncertainties.  

\begin{table}[ht]
\footnotesize
\begin{adjustwidth}{-.5in}{-.5in}
\setlength{\tabcolsep}{0.4em}
\centering
{\renewcommand{\arraystretch}{1.7}
\begin{tabular}{|c|c|c|c|c|c|c|c|c|}
\hline
\multirow{2}{*}{Cuts} & 
\multicolumn{3}{c|}{Signal $\big(m_{\ts}$, $m_{\X}\big)$ (GeV)} & 
\multirow{2}{*}{$t\bar{t}+W/Z$} &
\multirow{2}{*}{$t\bar{t}+j$} &
\multirow{2}{*}{single $t$} &
\multirow{2}{*}{$W/Z+j\!\!$} &
\multirow{2}{*}{QCD}
 \T\B\\ 
\hhline{~---~~~~}
 & (4000, 1) & (6000, 1) & (8000, 1) & & && &
\T\B
\\
\hline
$N_{\textrm{jet}} \ge$ 2 &
$\scinot{4.8}{3}$& $\scinot{5.3}{2}$& $\scinot{8.0}{1}$& $\scinot{1.6}{6}$& $\scinot{5.1}{7}$& $\scinot{5.4}{6}$& $\scinot{6.3}{7}$& $\scinot{2.8}{9}$
\T\B\\ \hline
$N_{\mu} \ge$ 1 &
$\scinot{9.1}{2}$& $\scinot{1.2}{2}$& $\scinot{2.1}{1}$& $\scinot{1.6}{5}$& $\scinot{4.3}{6}$& $\scinot{3.4}{5}$& $\scinot{5.3}{5}$& $\scinot{2.3}{7}$
\T\B\\ \hline
isolated $l^\pm$ veto &
$\scinot{9.1}{2}$& $\scinot{1.2}{2}$& $\scinot{2.1}{1}$& $\scinot{1.5}{5}$& $\scinot{4.1}{6}$& $\scinot{3.2}{5}$& $\scinot{5.3}{5}$& $\scinot{2.3}{7}$
\T\B\\ \hline
$\Delta\phi_{\small{\MET J}} >$ 1.0 &
$\scinot{5.0}{2}$& $\scinot{6.5}{1}$& $\scinot{1.2}{1}$& $\scinot{7.6}{3}$& $\scinot{1.6}{5}$& $\scinot{1.4}{4}$& $\scinot{3.3}{4}$& $\scinot{1.1}{6}$
\T\B\\ \hline
$\!\MET >$ 4.0 TeV$\!$ &
$\scinot{1.5}{1}$& $\scinot{1.7}{1}$& $7.2$& $2.9$& $\scinot{5.0}{-1}$& $\scinot{6.1}{-1}$& $\scinot{1.5}{-1}$& $\scinot{1.2}{-3}$
\T\B\\ \hline
\end{tabular}}
\caption{Background and signal yields for the heavy stops cut-flow in Section~\ref{sec:heavy_stop}, assuming 3000~fb$^{-1}$ of integrated luminosity. Single $t$ includes events with an extra $W/Z$.}
\label{table:cut-flow1}
\end{adjustwidth}
\end{table}

\subsection{Compressed Spectra}
\label{sec:compressed}
As the neutralino mass approaches the stop mass, both the $\MET$ and the top $p_T$ are reduced.  By relaxing some of the cuts in the previous section and trading the muon-in-jet requirement for an isolated lepton requirement, sensitivity to this region of parameter space can be improved.  Our cut-flow targeting the compressed region is:
\end{spacing}
\vspace{-2pt}
\begin{spacing}{1.15}
\begin{enumerate}
\item At least two anti-$k_T$ jets with cone parameter $\Delta R~=~0.5$.  The kinematic requirements $|\eta|<2.5$ and $p_{T}>500$ GeV are imposed. 
\vspace{-5pt}
\item Two isolated leptons (either electrons or muons) with $p_{Tl}>35$ GeV. A lepton satisfies the isolation cut when the total $p_{T}$ of all particles in a cone of $\Delta R=0.5$ around the lepton is less than 10\% of its $p_{T}$. 
\vspace{-5pt}
\item $\MET > 2$ TeV.
\vspace{-5pt}
\item $\Delta \phi_{\MET J,\,l} > 1.0$, where $\Delta \phi_{\MET J,\,l}$ is the smallest $|\Delta \phi|$ between $\MET$ and any jet with $p_{T}>200$ GeV and $|\eta|<2.5$, and any isolated lepton with $p_{Tl}>35$ GeV and $|\eta|<2.5$
\end{enumerate}
\end{spacing}
\begin{spacing}{1.3}
These requirements yield increased sensitivity for $m_{\ts} \lesssim 3$ TeV close to the diagonal of the $(m_{\ts}\,,\; m_{\X})$ plane.  
Table~\ref{table:cut-flow2} gives the results of this cut flow for the background and three signal mass points. Note that the $\MET > 2$ TeV requirement implicitly relies on the presence of extra QCD radiation in association with the signal.  This implies some uncertainty on initial-state radiation that we assume is covered by the systematic uncertainties applied on the signal samples.  Note that this cut-flow is much more sensitive to detector and machine details than the previous one.  We therefore present it only as a a proof of principle that going to higher energies does not necessarily imply sacrificing sensitivity to compressed, i.e. soft, physics. 

Including pile-up could have an important effect on the results for the compressed region.  An estimate for the energy deposited in a cone of radius $0.5$ at $\sqrt{s}=100$~TeV is $\approx 200~{\rm GeV}\left(\frac{L}{10^{34}\;{\rm cm}^{-2}\;{\rm s}^{-1}}\right)$~\cite{Barletta:2013ooa, Skands:2013asa}.  Most of this energy can be subtracted using common pile-up suppression techniques, so it is reasonable to expect small modifications to jet physics given the $p_T$ thresholds relevant for the models considered here.  The only possible exception is the $\Delta \phi$ requirement, which would be affected by resolution effects.  We verified that raising the jet $p_T$ threshold to $500$ GeV does not considerably impact our reach, giving us confidence that the impact of pile-up on the jet requirements will remain small.   However, lepton isolation may suffer more significantly, which would impact the results for the compressed scenarios.  Studies of such effects would require detailed assumptions of the detector performance, and thus we leave them for future work.

\begin{table}[ht]
\footnotesize
\begin{adjustwidth}{-.5in}{-.5in}
\setlength{\tabcolsep}{0.3em}
\centering
{\renewcommand{\arraystretch}{1.7}
\begin{tabular}{|c|c|c|c|c|c|c|c|c|}
\hline
\multirow{2}{*}{Cuts} & 
\multicolumn{3}{c|}{Signal $\big(m_{\ts}$, $m_{\X}\big)$ (GeV)} & 
\multirow{2}{*}{$t\bar{t}+W/Z$} &
\multirow{2}{*}{$t\bar{t}+j$} &
\multirow{2}{*}{single $t$} &
\multirow{2}{*}{$W/Z+j\!\!$} &
\multirow{2}{*}{QCD}
 \T\B\\ 
\hhline{~---~~~~~}
 & $\!(2000, 1500)\!$ & $\!(3000, 2500)\!$ & $\!(4000, 3500)\!$ & & && &
\T\B
\\
\hline
$N_{\textrm{jet}} \ge$ 2 & 
$\scinot{2.0}{5}$& $\scinot{2.6}{4}$& $\scinot{5.0}{3}$& $\scinot{3.9}{7}$& $\scinot{1.8}{9}$& $\scinot{2}{8}$& $\scinot{1.6}{9}$ & $\scinot{9.4}{10}$
\T\B\\ \hline
$N_\ell \ge$ 2 &
$\scinot{1.1}{3}$& $\scinot{1.6}{2}$& $\scinot{3.6}{1}$& $\scinot{4.1}{5}$& $\scinot{1.2}{7}$& $\scinot{1.1}{6}$& $\scinot{1.2}{4}$ & $\scinot{7.6}{1}$
\T\B\\ \hline
$|\Delta \phi_{\MET J,\,l}| > 1.0$ &
$\scinot{4.6}{2}$& $\scinot{7.1}{1}$& $\scinot{1.7}{1}$& $\scinot{4.1}{5}$& $\scinot{1.2}{6}$& $\scinot{1.1}{6}$& $\scinot{5}{2}$ & $0$
\T\B\\ \hline
$\!\MET >$ 2 TeV$\!$&
$6.8$& $5.3$& $2.9$& $1.2$& $\scinot{3.6}{-2}$& $\scinot{4.5}{-1}$& $0$ & $0$
\T\B\\ \hline
\end{tabular}}
\caption{Background and signal yields for the compressed spectra cut-flow in Section~\ref{sec:compressed}, assuming 3000~fb$^{-1}$ of integrated luminosity. Single $t$ includes events with an extra $W/Z$.}
\label{table:cut-flow2}
\end{adjustwidth}
\end{table}

\subsection{Results}
\label{sec:results}
\begin{figure}[ht]
\includegraphics[width=.47 \textwidth]{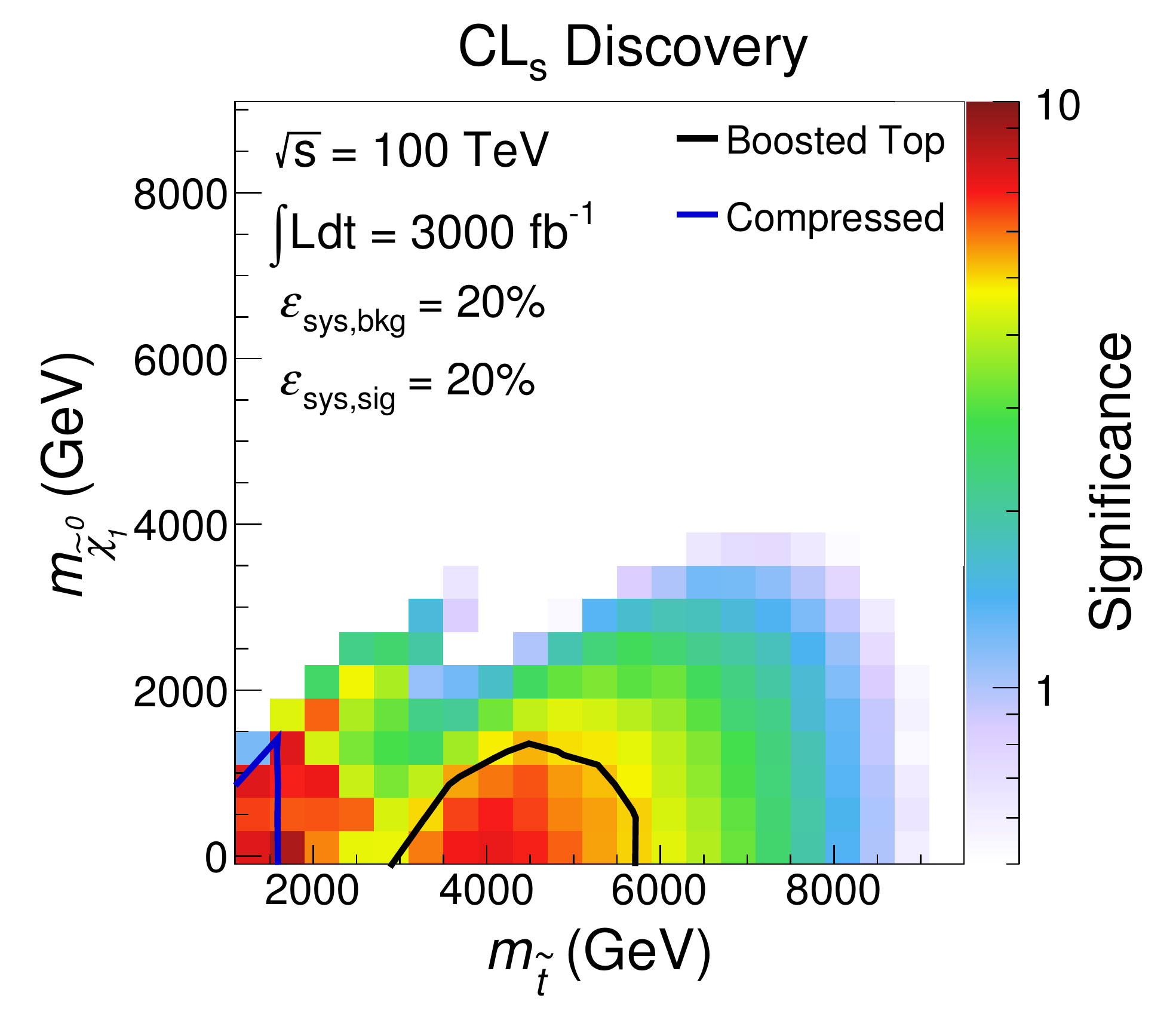}
\includegraphics[width=.50 \textwidth]{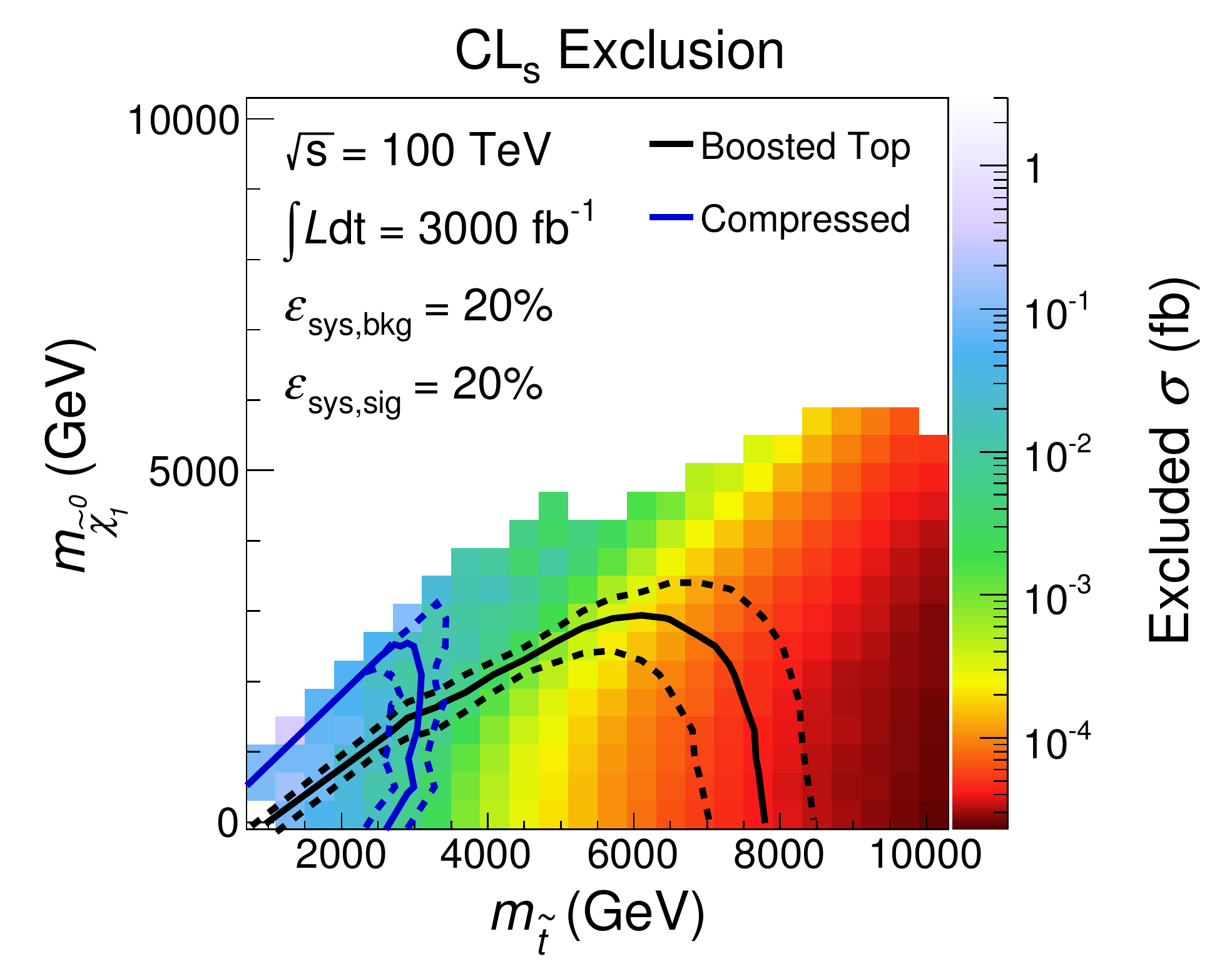}
\caption{Projected discovery potential [left] and exclusion limits [right] for 3000 fb$^{-1}$ of total integrated luminosity. At each signal point, the significance is obtained by taking the smaller $\textrm{CL}_s$ between the heavy stop and compressed spectra search strategies, and converting $\textrm{CL}_s$ to number of $\sigma$'s. The blue and black contours (dotted) are the expected ($\pm 1 \sigma$) exclusions/discovery contours using the heavy stop and compressed spectra searches.  }
\label{fig:exclusion_discovery}
\end{figure}

Figure~\ref{fig:exclusion_discovery} shows the exclusion and discovery potential utilizing the cut-flows discussed in the previous section. Results are presented in the stop-neutralino mass plane assuming systematic uncertainties of $20\%$ on the background and signal yields. The discovery potential and mass reach are shown in Sec.~\ref{sec:lumi}-\ref{sec:sys} for different choices of integrated luminosities and systematic uncertainties.

The exclusion is obtained using $\textrm{CL}_s$ statistics, where the background and signal are modeled as Poisson distributions.   A signal point is rejected for $\textrm{CL}_s < 0.05$. Alternatively, a signal is discovered when the $\textrm{CL}_s$ for the background only hypothesis is less than $\sim \scinot{3}{-7}$, corresponding to 5$\sigma$. The expected exclusion limits and $\pm 1\sigma$ contours are computed using {\tt ROOSTATS}~\cite{Moneta:2010pm}.

Stops with masses up to $\approx 5.5$ TeV can be discovered when the neutralino is massless, assuming $3000\;{\rm fb}^{-1}$ of integrated luminosity.  The exclusion reach is $\approx 8$ TeV, which corresponds to $\sim 100$ signal events before cuts.  Note that this agrees with the estimate obtained by extrapolating the number of excluded signal events at $\sqrt{s}=8$~TeV~\cite{CReach}.  Since we optimized for exclusion as opposed to discovery, there is a gap between the discovery contours of the two different search strategies.

The searches proposed here also have good discriminating power away from the massless neutralino limit.  A $1.5$ TeV stop could be discovered in the compressed region of parameter space.  It is possible to exclude neutralino masses up to $2$ TeV in most of the parameter space. 

All of the results presented here have been obtained with very minimal cut-flows that do not rely on $b$-tagging or jet substructure techniques.  Additional refinements should increase the search sensitivity, at the price of making assumptions on the future detector design.  

\subsection{Different Luminosities}
\label{sec:lumi}
An open question in the design for the 100~TeV proton-proton collider is the luminosity that is necessary to take full advantage of the high center of mass energy.  As cross sections fall with increased center of mass energy, one should expect that higher energy colliders require more integrated luminosity to fulfill their potential.  The necessary luminosity typically scales quadratically with the center of mass energy, meaning that one should expect that the 100~TeV proton-proton collider would need roughly 50 times the luminosity of the LHC at 14~TeV.

This section shows the scaling of our search strategy as a function of the number of collected events.  As the luminosity changes, we re-optimize the $\MET$ cut.
For integrated luminosities of $300\ifb$, a $\MET$ cut of $3$ TeV is chosen. For $30000\ifb$, a $\MET$ cut of $5$ or $6$ TeV is chosen, depending on the mass point. Table~\ref{table:bkg_lumi} lists the number of background events for the heavy stop search and these two choices of luminosity and $\MET$ cut.  Figure~\ref{fig:vary_lumi_300} (\ref{fig:vary_lumi_30000}) shows the expected $\textrm{CL}_s$ discovery and exclusion for 300 (30000)$\ifb$ of integrated luminosity. For $300\ifb$, the discovery potential is limited, but we obtain a $3 \sigma$ evidence in the bulk of the parameter space. With $30000\ifb$, stops of 8 (10) TeV could be discovered (excluded), a clear improvement over the $3000\ifb$ result.

Assuming a constant systematics of 20\% for both signal and background, if we model the mass reach as a function of luminosity as 
\be
\frac{1}{n(\LL) }= \frac{d \log m_{\tilde{t}~2\sigma}(\LL)}{ d \log \LL}
\ee
then we find $n\simeq 7$ in the $300 \ifb$ to $3000\ifb$ range of luminosities and $n\simeq 10$ in the $3000\ifb$ to $30000\ifb$ range. This indicates that the 100~TeV collider is still gaining significant reach at $3000\ifb$ and running out reach at $30000\ifb$. Reaching a higher integrated luminosity implies running at higher instantaneous luminosity, with potential implications for detector performance that we do not consider here.  However, we expect that improvements in detector design and pile-up mitigation strategies will minimize any loss of sensitivity from harsher running conditions.

\begin{figure}[ht!]
\includegraphics[width=.46 \textwidth]{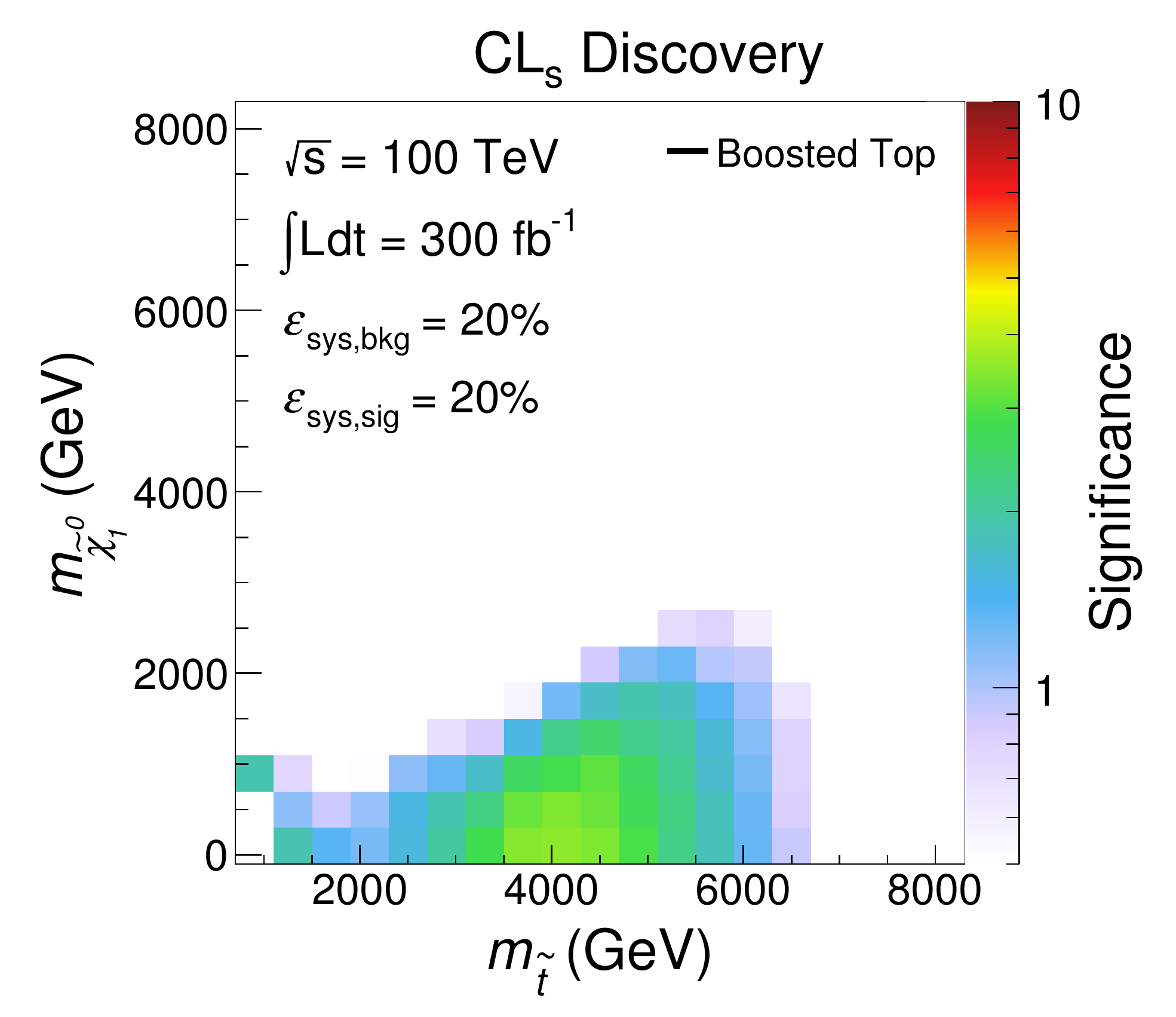}
\includegraphics[width=.50 \textwidth]{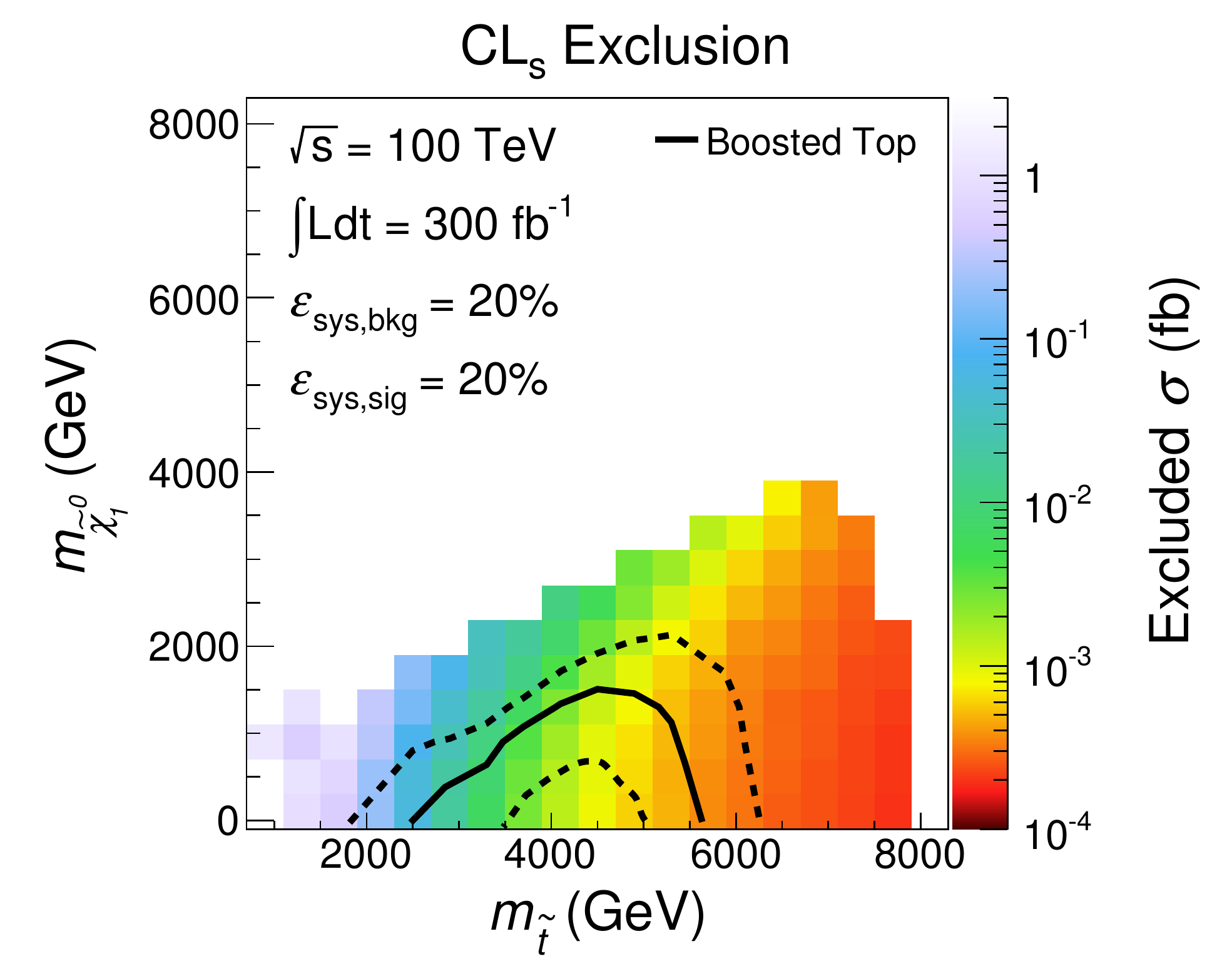}
\caption{Discovery [left] and exclusion [right] limits with an integrated luminosity of 300 fb$^{-1}$. Only the heavy stop search is shown.}
\label{fig:vary_lumi_300}
\end{figure}

\begin{figure}[ht!]
\includegraphics[width=.46 \textwidth]{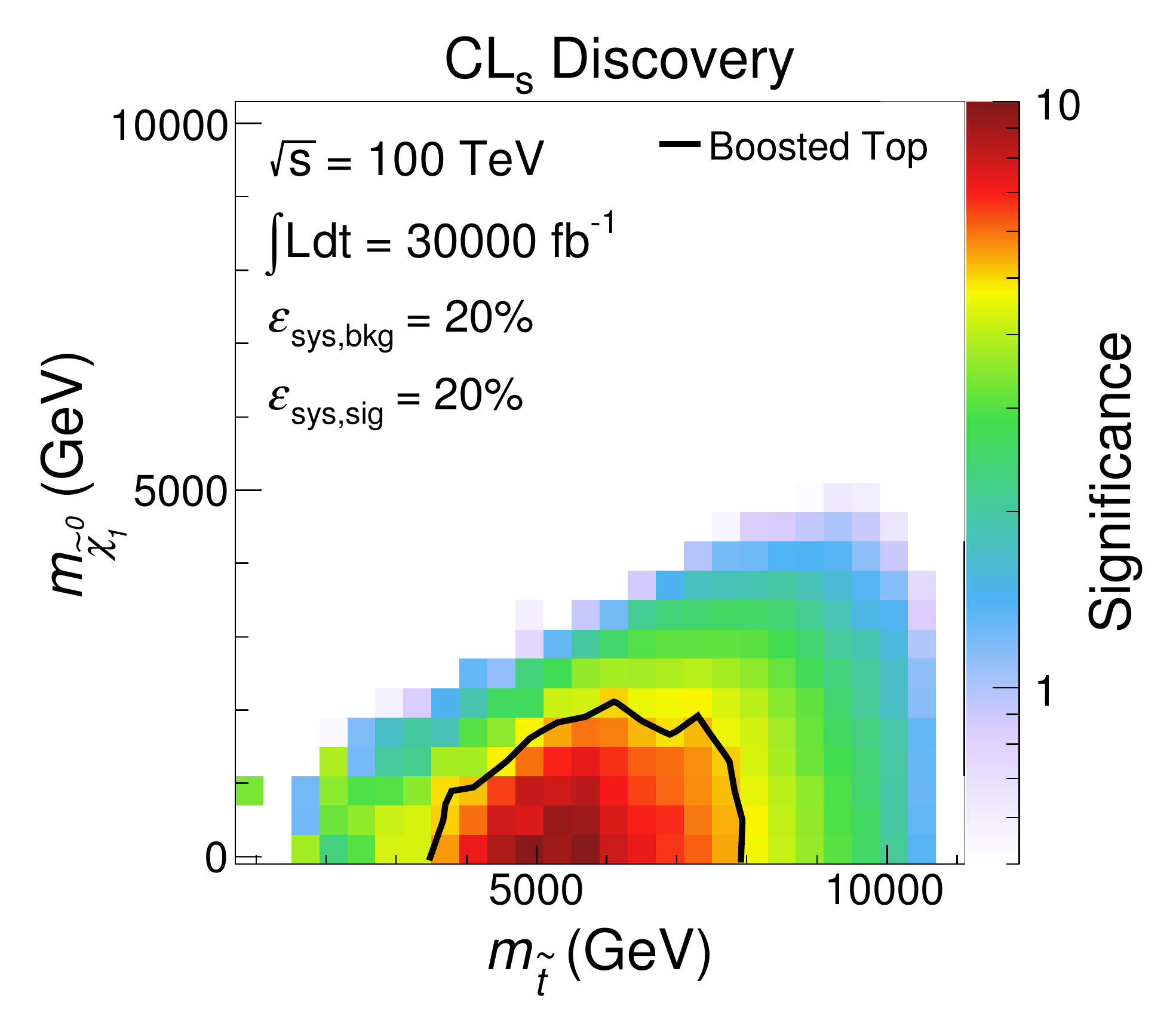}
\includegraphics[width=.50 \textwidth]{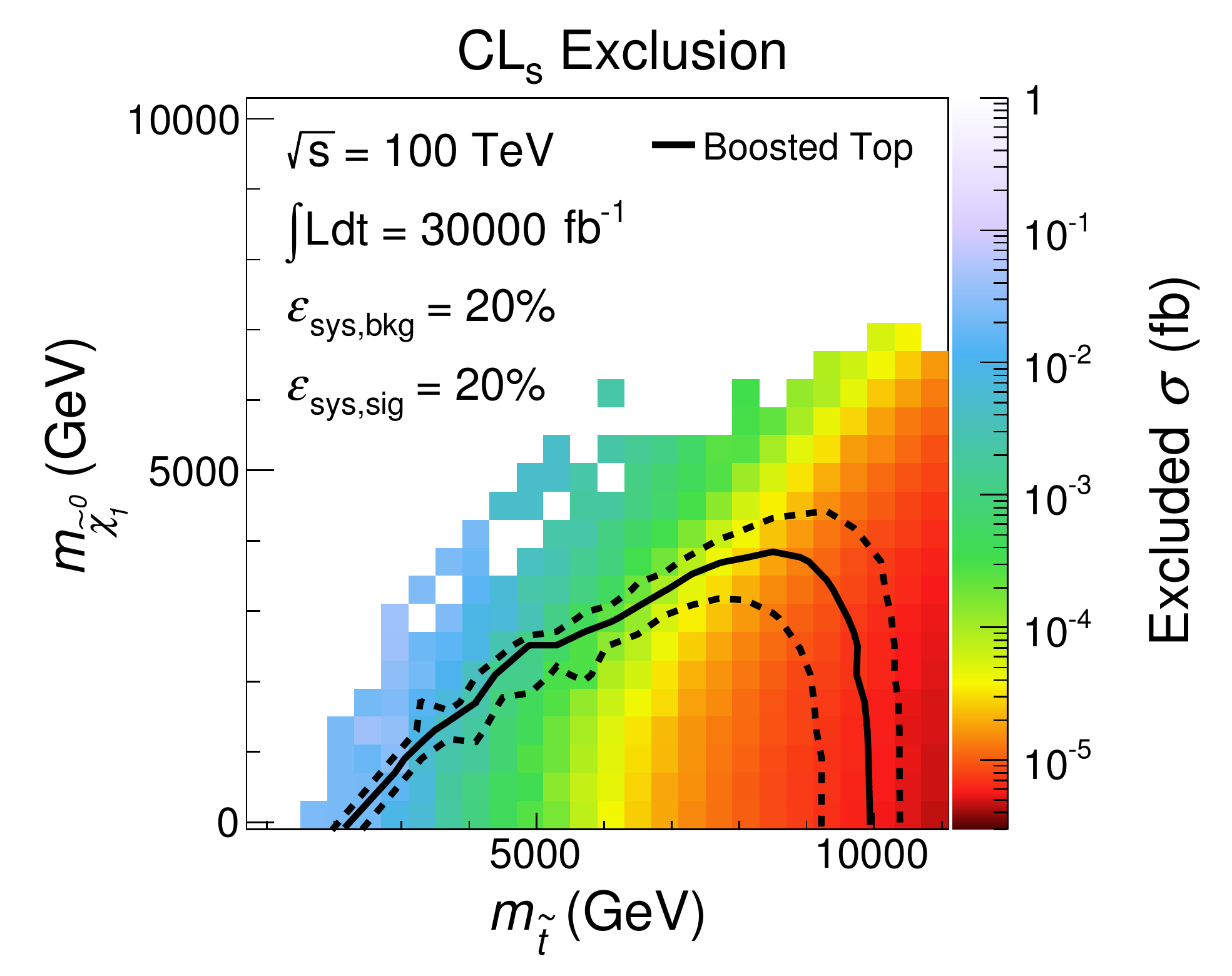}
\caption{Discovery [left] and exclusion [right] limits with an integrated luminosity of 30000 fb$^{-1}$. Only the heavy stop search is shown.}
\label{fig:vary_lumi_30000}
\end{figure}

\begin{table}[ht!]
\footnotesize
\begin{adjustwidth}{-.5in}{-.5in}
\setlength{\tabcolsep}{0.3em}
\centering
{\renewcommand{\arraystretch}{1.7}
\begin{tabular}{|c|c|c|c|c|c|c|c|c|}
\hline
\multirow{2}{*}
{\parbox[t]{3.5cm}{Luminosity (fb$^{-1}$),\\
 $\MET$ cut (GeV)}} & 
\multicolumn{3}{c|}{Signal $\big(m_{\ts}$, $m_{\X}\big)$ (GeV)} & 
\multirow{2}{*}{$t\bar{t}+W/Z$} &
\multirow{2}{*}{$t\bar{t}+j$} &
\multirow{2}{*}{single $t$} &
\multirow{2}{*}{$W/Z+j\!\!$}&
\multirow{2}{*}{QCD}
 \T\B\\ 
\hhline{~---~~~~~}
 & $\!(6000, 1)\!$ & $\!(8000, 1)\!$ & $\!(10000, 1)\!$ & & && &
\T\B
\\
\hline
$300,3000$ & 
$4.5$& $1.0$& $\scinot{2.2}{-1}$& $1.3$& $\scinot{3.1}{-1}$& $\scinot{3.5}{-1}$& $\scinot{1.5}{-2}$& $\scinot{1.9}{-4}$
\T\B\\ \hline
$3000,4000$ &
$\scinot{2.1}{1}$& $7.2$& $1.8$& $3.4$& $\scinot{5}{-1}$& $\scinot{6.1}{-1}$& $\scinot{1.5}{-1}$& $\scinot{1.2}{-3}$
\T\B\\ \hline
$30000,6000$&
$\scinot{1.6}{1}$& $\scinot{2.3}{1}$& $9.3$& $4.3$& $\scinot{1.2}{-1}$& $\scinot{3.8}{-1}$& $0$ & $\scinot{6.9}{-3}$
\T\B\\ \hline
\end{tabular}}
\caption{Background and signal yields for three different choices of luminosity and three different heavy stop search signal regions. Single $t$ includes events with an extra $W/Z$.}
\label{table:bkg_lumi}
\end{adjustwidth}
\end{table}

\subsection{Different Systematics}
\label{sec:sys}
This section explores how the exclusion and discovery potential changes as a function of systematic uncertainty.  For the results in Sec.~\ref{sec:results} a systematic uncertainty of $\sys=20\%$ for both background and signal was assumed. The signal regions proposed in this paper yield $\mathcal{O}(5)$ events for both background and signal when the masses are chosen at the edge of the exclusion reach.  
Figure~\ref{fig:vary_systematics} illustrates how the exclusion changes as the signal (background) uncertainty in the left (right) panel is increased from $20\%$ to $50\%$.  The exclusion is robust against changes in background systematics.  A change in signal uncertainty results in a modest shift of the limits, since the signal hypothesis becomes harder to exclude when marginalized over larger systematics.

\begin{figure}[ht]
\includegraphics[width=.48 \textwidth]{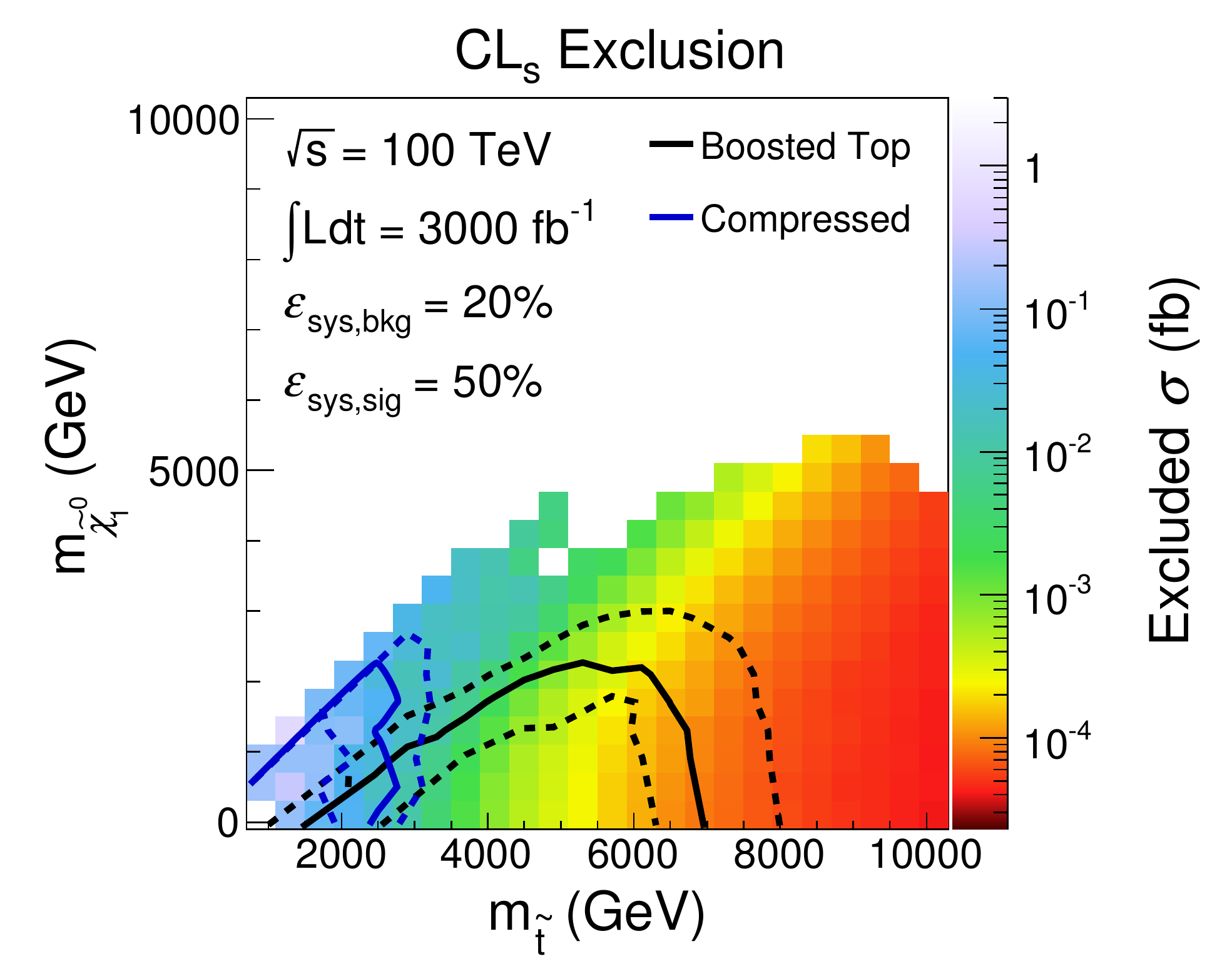}
\includegraphics[width=.48 \textwidth]{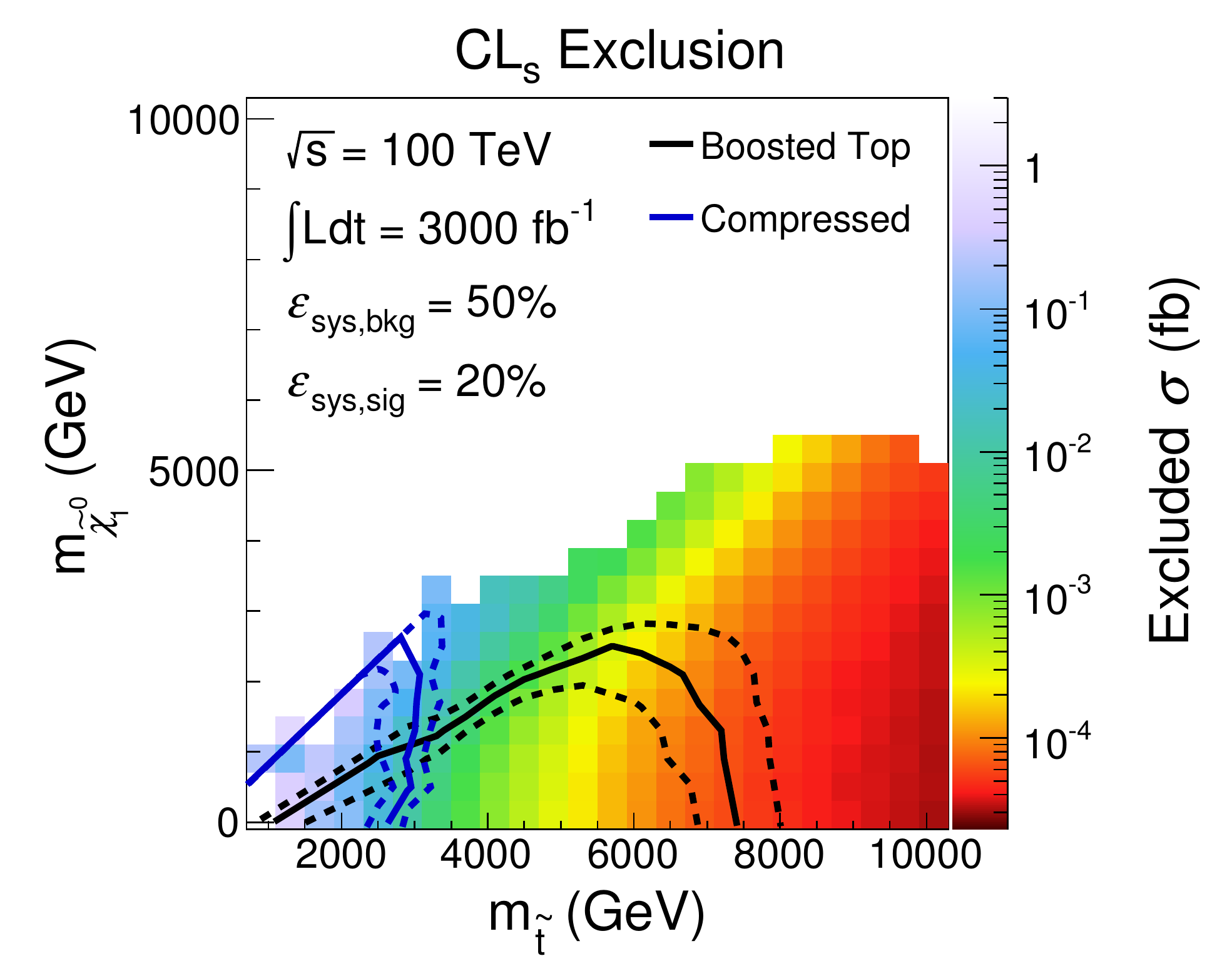}
\caption{Expected exclusion limit with background and signal errors fixed to $(\varepsilon_{\textrm{sys,bkg}}, \varepsilon_{\textrm{sys,sig}}) = (20\%, 50\%)$ [left].   Expected exclusion limit with $(\varepsilon_{\textrm{sys,bkg}}, \varepsilon_{\textrm{sys,sig}}) = (50\%, 20\%)$ [right].}\label{fig:vary_systematics}
\end{figure}

Figure~\ref{fig:vary_systematics} shows the same for the discovery potential. The expected discovery changes modestly as the systematic uncertainty  on the signal is increased.  However, when the background systematic uncertainty is increased to $50\%$, discovery becomes impossible with 3000 fb$^{-1}$; only $3\sigma$ evidence is possible in the bulk of the parameter space.  A larger background systematic uncertainty implies that it is harder to reject the background hypothesis, so a precise understanding of the backgrounds will be crucial for discovery.

\begin{figure}[ht]
  \includegraphics[width=.48 \textwidth]{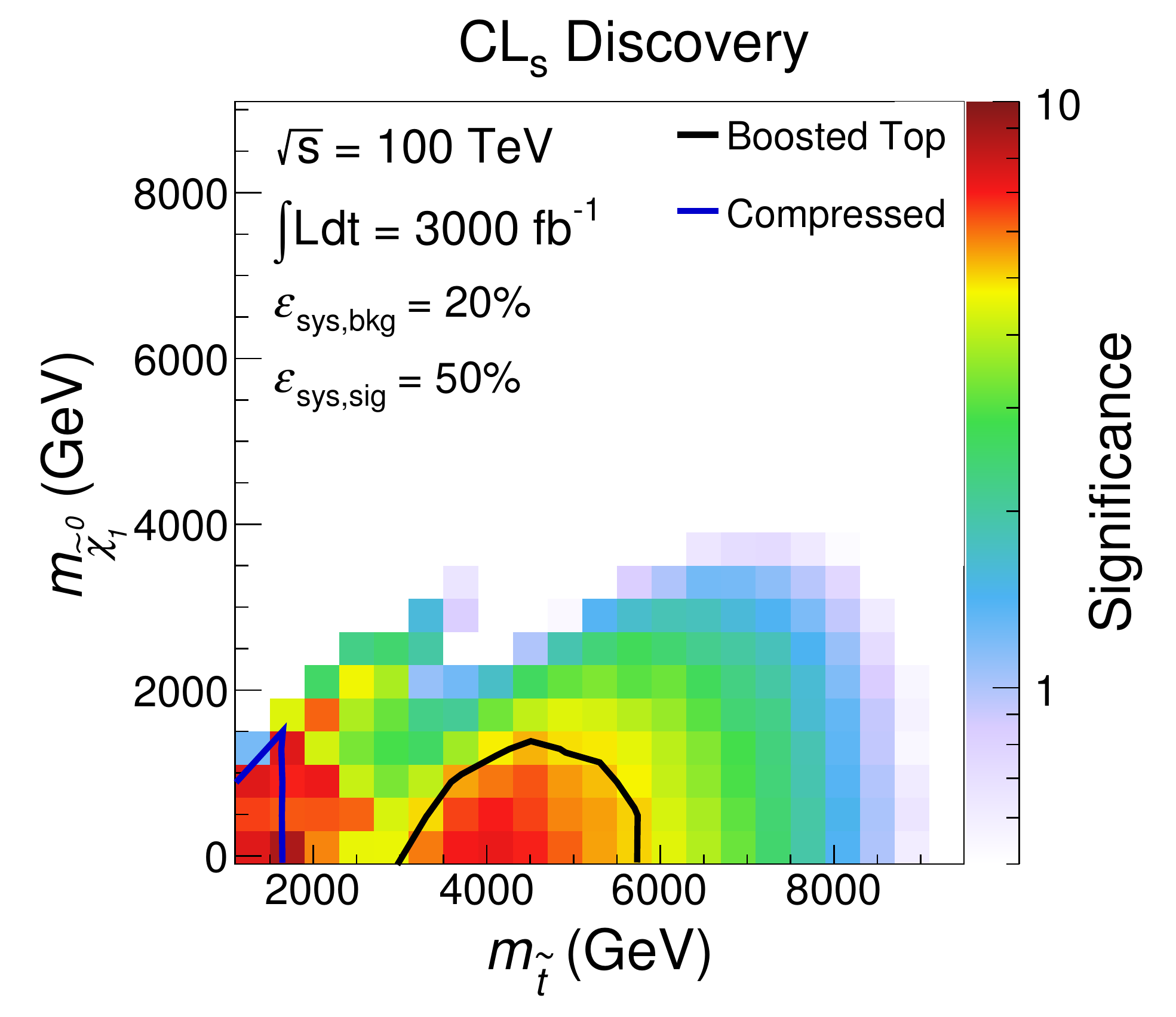}
  \includegraphics[width=.48 \textwidth]{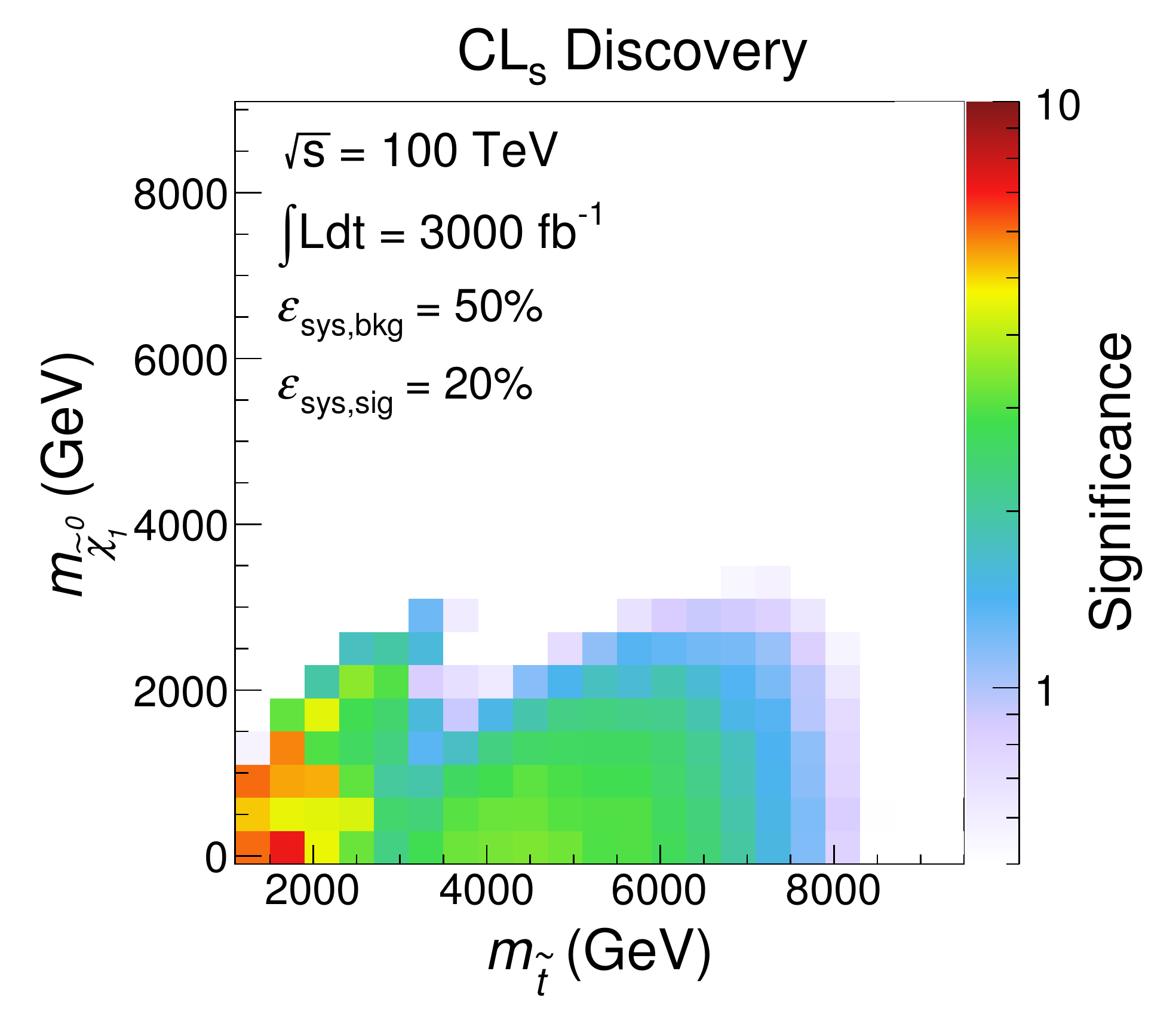}
\caption{Discovery potential with background and signal errors fixed to $(\varepsilon_{\textrm{sys,bkg}}, \varepsilon_{\textrm{sys,sig}}) = (20\%, 50\%)$ [left].   Discovery potential with systematics $(\varepsilon_{\textrm{sys,bkg}}, \varepsilon_{\textrm{sys,sig}}) = (50\%, 20\%)$ [right].}
\label{fig:vary_systematics_dis}
\end{figure}

\pagebreak

\section{Conclusions}
\label{sec:conclusion}
In this paper we propose a robust search strategy targeting stops that decay to a top quark and a stable neutral particle at a 100~TeV proton-proton collider.  A 5.5 (8) TeV stop could be discovered (excluded) at such a machine with 3000 $\ifb$ of integrated luminosity.

Such an exclusion would have a deep impact on our understanding of electroweak fine-tuning.  In the Minimal SUSY SM (MSSM), the tuning of the electroweak scale, $\Delta^{-1}$, receives a large contribution from the SUSY breaking parameters in the stop sector.  A rough estimate of the minimum contribution to the Higgs mass parameter yields~\cite{Papucci:2011wy}:
\be
\label{eq:tuning}
\left(\frac{\Delta^{-1}}{2\times 10^{-4}}\right)\approx \left(\frac{10\; \TeV}{m_{\widetilde{t}}}\right)^{2}\sin^2\beta \left(\frac{\log(\Lambda/\TeV)}{5}\right)^{-1}\, , \\
\ee
where $\Lambda$ is the SUSY breaking scale and $\tan\beta=v_u/v_d$ the ratio of the two Higgs doublets vacuum expectation values.
A 100 TeV proton collider clearly has the potential to impact our understanding of electroweak naturalness to an unprecedented degree.

However stop masses approaching 10 TeV are above the typical range motivated by fine-tuning considerations. Nonetheless, this range of masses could be the consequence of the Higgs mass being so far above $m_Z$.  In the MSSM, the Higgs quartic coupling must receive sizable radiative corrections to raise the Higgs boson mass from $m_{Z}$ to the observed value.  The largest of these contributions arise from the top sector and comes in two forms.  In the effective theory below the stop mass, the first is the contribution from the top quark and is logarithmically enhanced by the running from the mass of the top squark down to the top quark.  The second contribution is given by the $A$-terms which at low energies can be viewed as finite threshold corrections.  In order to have top squarks with masses in the range accessible by the  LHC14, there need to be sizable $A$-terms. However, many calculable frameworks for coupling a SUSY breaking sector to the visible sector result in suppressed $A$-terms, \emph{e.g.} gauge mediation \cite{Giudice:1998bp}, anomaly mediation \cite{Giudice:1998xp, Randall:1998uk}, and gaugino mediation \cite{Chacko:1999mi}.  These classes of theories are the ones that have the best solutions for the SUSY flavor problem and hence are amongst the most favored.   In the absence of sizable $A$-terms, only the logarithmically enhanced top quark contributions are left to raise the Higgs mass which results in top squarks with masses in the range $6 \text{ TeV} \lesssim m_{\widetilde{t}} \lesssim 10 \text{ TeV}$ at the 2$\sigma$ level (for large $\tan \beta$ and small values of $\mu$) \cite{Draper:2013oza}.  These masses are outside the reach of the LHC14, but discoverable at a 100 TeV collider.  Frequently the top squarks are amongst the lighter colored superpartners, meaning that it is possible that supersymmetry will be above the reach of the LHC14.  This observation provides motivation for building another energy frontier machine, even in the case where no new physics is found at the LHC14.  

Beyond the theory motivation, the lessons of this study can be generalized to a wide class of searches for boosted top quarks signatures.  In particular, there are important implications for future detector design. For example, a granularity of $\Delta \phi \times \Delta \eta \approx 0.02\times 0.02$ is needed if hadronic substructure techniques are going to be effective.  This requirement might be relaxed by relying on tracking information incorporated into a more complicated reconstruction algorithm such as Particle Flow. On the other hand, requiring a muon within a jet is a simple and robust way to exploit the qualitative differences between new physics and SM backgrounds that does not require detector improvements beyond what the LHC can do today.  

Furthermore we have shown that it would be desirable to achieve higher integrated luminosity than presently used in 100 TeV studies. The current benchmark of $3000\ifb$~\cite{Cohen:2013xda, Andeen:2013zca, Apanasevich:2013cta, Degrande:2013yda, Stolarski:2013msa, Yu:2013wta, Zhou:2013raa, Low:2014cba} does not saturate the physics reach of this machine. The ideal integrated luminosity would be $10000-30000\ifb$.

Designing searches for heavy stops yields a concrete example of how a 100 TeV collider is qualitatively different from the LHC. The new energy regime that this machine will explore is so far above the electroweak scale as to render traditional search strategies ineffective.  On the other hand, this makes the analyst's job easier since signals and backgrounds become more qualitatively different.  This is exemplified by the sensitivity that can be derived using the simple cut-flows presented in this work.

\section*{Acknowledgements}
We thank Nima Arkani-Hamed, Andy Haas, Stephanie Majewski, Gigi Rolandi, Maurizio Pierini, Michele Selvaggi, Anyes Taffard, Daniel Whiteson. We thank Christoph Borschensky, Michael Kranmer and Tilman Plehn for providing the NLL + NLO stop production cross sections. TC  is supported by DoE contract number DE-AC02-76SF00515 and in part by the NSF grant NSF PHY11-25915.  RTD is supported by the NSF grant PHY-0907744.  MH is supported by DoE contract number DE-AC02-05CH11231.  HKL is supported by the DOE SCG Fellowship.  

\end{spacing}
\pagebreak
\begin{spacing}{1.1}
\bibliography{StopsAt100TeV}
\bibliographystyle{utphys}
\end{spacing}
\end{document}